\documentclass[aps,prc,reprint,showpacs]{revtex4-1}
\usepackage{amsmath}
\usepackage{bm}
\usepackage{graphicx}
\usepackage{hyperref}
\usepackage{subfigure}
\usepackage{float}

\begin{document}

%\preprint{}

\title{Rapidity Correlation Structure in Nuclear Collisions}%from Causal Hydrodynamics}

\author{Sean Gavin,$^1$ George Moschelli,$^2$ and Christopher Zin$^1$} 
\address{
$^1$Department of Physics and Astronomy, Wayne State University, 
Detroit, MI, 48202\\
$^2$Lawrence Technological University, 21000 West Ten Mile Road, Southfield, MI  48075}

%
%\author{Sean Gavin}
%\email{sean.gavin@wayne.edu}
%\affiliation{Department of Physics and Astronomy, Wayne State University, 666 W Hancock, Detroit, MI, 48202, USA}
%%
%\author{George Moschelli}
%\email{gmoschelli@ltu.edu}
%\affiliation{Lawrence Technological University, 21000 West Ten Mile Road, Southfield, MI  48075-1058}
%%
%\author{Rajendra Pokharel}
%\email{rpokharel@ltu.edu}
%\affiliation{Lawrence Technological University, 21000 West Ten Mile Road, Southfield, MI  48075-1058}


\date{\today}

\begin{abstract}
We show that measurements of the rapidity dependence of transverse momentum correlations can be used to determine the characteristic time $\tau_\pi$ that dictates the rate of isotropization of the stress energy tensor, as well as the shear viscosity $\nu = \eta/sT$. We formulate methods for computing these correlations using second order dissipative hydrodynamics with noise. Current data are consistent with $\tau_\pi/\nu \sim 10$, but targeted measurements can improve this precision.  
\end{abstract}

%\begin{abstract}
%The collective flow exhibited by high energy nuclear collision experiments is widely attributed to hydrodynamic flow. We propose that measurements of the rapidity dependence of transverse momentum correlations can betray the onset of hydrodynamics. To compute the necessary correlation functions, we formulate methods using second order dissipative hydrodynamics with noise. We show how measurements can be used to determine both the shear viscosity $\nu = \eta/sT$ and -- for the first time -- the characteristic time $\tau_\pi$ that dictates the rate of isotropization of the stress energy tensor. Current data are consistent with $\tau_\pi/\nu \sim 10$, but we argue that refined measurements can improve this precision.  
%\end{abstract}

%\pacs{}
%\keywords{}

\maketitle

%		INTRODUCTION
%
\section{\label{sec:intro} Introduction}

The forces that drive the nuclear collision system towards local thermal equilibrium leave few observable traces.  
%The forces that drive the high density partonic system produced by a nuclear collision towards local thermal equilibrium leave few experimental traces.  
 Heavy ion experiments report a range of features widely attributed to the hydrodynamic flow of a near-equilibrium quark gluon plasma at the Relativistic Heavy Ion Collider, RHIC, and the Large Hadron Collider, LHC. In particular, measurements of azimuthal anisotropy provide the most comprehensive support for the hydrodynamic description of these systems \cite{Shen:2015msa}. In search of the source of this flow, experimenters turned to smaller proton-proton, proton-nucleus, and deuterium-nucleus collisions, expecting to find this effect absent. Instead, these collisions show an  azimuthal anisotropy that is comparable to the larger ion-ion systems \cite{Khachatryan:2010gv,Chatrchyan:2013nka,Aad:2013fja,Abelev:2012ola,Adare:2013piz}. How can we learn about the mechanisms that give rise to hydrodynamics if every available collision system exhibits flow? 

In this paper we argue that the rapidity dependence of transverse momentum correlations can be used to extract information on the thermalization process.   In ref.\ \cite{Gavin:2006xd} we pointed out that viscous diffusion broadens the rapidity dependence of $p_t$ correlations, yielding information on the shear viscosity. Here we propose that systematic changes in the {\em shape} of this rapidity dependence with centrality can be used to measure $\tau_\pi$, the relaxation time  that sets the rate at which the pressure becomes isotropic.  

%How can we learn about the onset of hydrodynamics if every available collision system exhibits similar flow? In this paper we argue that the rapidity dependence of transverse momentum correlations can be used to extract information on the character of thermalization. Two particle correlations show a near-side peak that sits atop a ridge in relative rapidity; see e.g., \cite{Wenger:2008ts,Alver:2009id,Daugherity:2008su,Adare:2008ae,Adam:2015gda,Chatrchyan:2012wg,Khachatryan:2016ibd,Aad:2012bu,Wang:2013qca} and references cited therein.  In ref.\ \cite{Gavin:2006xd} we argued that the width of this peak in rapidity for $p_t$-weighted correlations bear information on the shear viscosity.  Here we propose that systematic changes in {\em shape} of the peak with centrality can be used to measure the relaxation time $\tau_\pi$ that sets rate at which the pressure becomes isotropic.  

Rapidity correlations provide the space-time information that allows us to probe the onset of hydrodynamic behavior in collisions. Two particle correlation measurements show a near-side peak that sits atop a flat ridge in relative rapidity; see, e.g., \cite{Wenger:2008ts,Alver:2009id,Daugherity:2008su,Adare:2008ae,Adam:2015gda,Chatrchyan:2012wg,Khachatryan:2016ibd,Aad:2012bu,Wang:2013qca}. This result affirms the long-standing principle that longitudinal expansion roughly follows a one-dimensional Hubble-like behavior \cite{Bjorken:1976mk,Shuryak:1978ij,Bjorken:1982qr}.  Long range correlations over several rapidity units originate at  the  earliest stages of an ion collision \cite{Dumitru:2008wn,Gavin:2008ev}. Correlated particles that are closer in rapidity interact for a longer period, depending on their rapidity separation.  Here, we are most interested in the short range behavior -- the peak -- because it tells us how fluctuations are dissipated by the stochastic dynamics of the strongly interacting system. 

We focus on transverse momentum correlations because they are dissipated by shear viscous diffusion, which is particularly sensitive to the relaxation time $\tau_\pi$. %That $p_t$ fluctuations are dissipated  by shear viscosity was first pointed out in ref.\ 
%We focus on transverse momentum correlations because they are dissipated by viscous diffusion \cite{Gavin:2006xd}. 
Nuclear collisions produce a fluid that flows with a transverse velocity that differs slightly from point to point within each event.  Viscous friction arises as neighboring fluid elements flow past one another. As a simple illustration, consider the variation of the velocity $v_x$ along the longitudinal $z$ direction near a point where the fluid is at rest. Near local equilibrium, this flow produces a stress 
\begin{equation}
S_{zx} = -\eta \partial v_x/\partial z 
\label{eq:NavSto}
\end{equation}
%
%in the Navier-Stokes regime  
that works to make the transverse momentum distribution as uniform as possible.  In refs.\ \cite{Gavin:2006xd,Pratt:2010zn}, we studied how transverse momentum fluctuations are spread throughout the liquid by viscous diffusion described by (\ref{eq:DiffModes}) in sec.\ \ref{sec:Modes}, which follows from the Navier-Stokes equation.  The effective diffusion coefficient is the kinematic viscosity $\nu = \eta/sT$, where $s$ is the entropy density and $T$ is the temperature. 

The character of viscous diffusion changes dramatically as the system evolves from its initial state toward the Navier-Stokes regime. In essence, the stress energy tensor $T_{zx}$ relaxes to (\ref{eq:NavSto}) at a rate 
\begin{equation}
\frac{\partial}{\partial t}T_{zx} = -\frac{1}{\tau_\pi}(T_{zx} -  S_{zx}),
\label{eq:IsStew}
\end{equation}
with corrections to be discussed later.  Now described by (\ref{eq:2DiffModes}), relaxation allows shear stress to propagate as waves. % for times $< \tau_\pi$.  
 In the next sections we will argue that the rapid longitudinal expansion in nuclear collisions can freeze a wavelike structure into the rapidity distribution, allowing an experimental glimpse of the equilibration process and a measurement of $\tau_\pi$.   
%
%Several experiments have measured transverse momentum fluctuations at RHIC and LHC, but so far only STAR has measured their pseudoraipidity dependence.  
%
Measurements from the STAR collaboration at RHIC discussed in secs.\ \ref{sec:constNu} and \ref{sec:betaPlots} may hint at these effects \cite{Agakishiev:2011fs,PrivComm}.  

We comment that several time scales describe different aspects of hydrodynamization. Near local equilibrium the relaxation times for shear, bulk, and heat transport are $\tau_\pi$, $\tau_\Pi$, and $\tau_q$, respectively \cite{Betz:2008me,Romatschke:2009im,Betz:2010cx,Denicol:2012es,Denicol:2014vaa}. At sufficiently low densities, these quantities can be calculated from the Boltzmann equation. However, causality arguments suggest that the general form of (\ref{eq:IsStew}) and similar bulk and heat relaxation equations can apply more widely \cite{1963AnPhy..24..419K,forster1995hydrodynamic}. Color fields evolve with their own distinct time scales \cite{Mrowczynski:2016etf}. Sometimes we describe the entire evolution from free-streaming partons to thermalized matter using a Boltzmann equation with a single effective relaxation time. Such time scales may differ appreciably.  Furthermore, pre-equilibrium flow affects all hydrodynamic observables. %The iconic pre-equilibrium probes are  dilepton, photon, and heavy quark production. 
Several groups have recently studied the effect of anisotropic pressure on single particle spectra, flow harmonics, and other bulk observables \cite{Broniowski:2008qk,Wu:2011yd,Heller:2012km,Gelis:2013rba,vanderSchee:2013pia,Liu:2015nwa}. It is unlikely that any single signal will yield unambiguous information on pre-equilibrium evolution. All of these signals together will likely be needed to create a complete picture.  

This paper is organized as follows.  We will discuss the rapidity dependence of transverse momentum fluctuations in terms of shear hydrodynamic modes. In sec.\ \ref{sec:Modes} we briefly introduce the hydrodynamic modes in the first and second order theory.  As pointed out in refs.\ \cite{Aziz:2004qu,Gavin:2006xd}, to discuss the evolution of these fluctuating modes towards the proper local equilibrium state, we must include hydrodynamic noise \cite{Calzetta:1997aj,Kapusta:2011gt,Kumar:2013twa}.  In sec.\ \ref{sec:Roadmap} we use the stochastic differential equations that describe hydrodynamics in the presence of noise to obtain equations for correlation functions.  We use analytic techniques for working with stochastic differential equations that are common in mathematics but less familiar in physics \cite{van2011stochastic,gardiner2004handbook}.  Our main result (\ref{eq:AbdelAzizEquation}) describing correlations in second order hydrodynamics is new. In sec.\ \ref{sec:ionCollisions} we derive equations that describe the fluctuations in the presence of Bjorken flow, in particular (\ref{eq:2VisDiff}). Both secs.\  \ref{sec:Roadmap} and \ref{sec:ionCollisions} are technical and may be skipped by readers interested only in the phenomenology. 

We then turn to the phenomenological problem at hand in sec.\ \ref{sec:Observables}, where we discuss the observables that we recommend to study correlations in collisions following refs.\ \cite{Gavin:2006xd,Pratt:2010zn}. In sec.\ \ref{sec:constNu} we solve (\ref{eq:2VisDiff}) and compare the results to data and to first order diffusion theory \cite{Gavin:2006xd}.  We find that data are best described by the second order theory \cite{Agakishiev:2011fs,PrivComm}.  Finally, we discuss how to measure $\tau_\pi$ in sec.\ \ref{sec:betaPlots}.

\section{Shear Hydrodynamic Modes}\label{sec:Modes}
Our description of the rapidity dependence of $p_t$ correlations begins with the observation that the spread of transverse velocity fluctuations in the beam direction is determined by shear hydrodynamic modes \cite{Gavin:2006xd}. Shear modes generally account for the linear response of a fluid in directions perpendicular to an initial impulse. Viscous diffusion spreads this response throughout the fluid, eventually making the velocity as uniform as possible. 
While shear modes likely dominate the observables we discuss here, other hydrodynamic modes exist and contribute elsewhere. Sound modes are compression waves that propagate in the same direction as the initial impulse. Additional diffusive modes transport conserved charges relative to energy density. Net charge and baryon number correlations in rapidity can be used to extract experimental information about diffusion coefficients \cite{Aziz:2004qu}. 

To identify the hydrodynamic modes, we consider fluctuations of a fluid at rest with energy density $e$ and pressure $p$. Small fluctuations produce a small velocity $\mathbf{v}$ corresponding to a momentum current $\mathbf{M} \approx (e+p)\mathbf{v}$. 
%%%%%%%%%
To linear order in the fluctuations, we write the conservation form of the relativistic Navier-Stokes equation:
\begin{equation}\label{eq:NavStokes2}
 {{\partial }\over{\partial t}} \mathbf{M}   + \bm{\nabla} p = {{\zeta + \tfrac{1}{3}\eta}\over{w}}\bm{\nabla}(\nabla\cdot \mathbf{M}) + \frac{\eta}{w}\nabla^2\mathbf{M},
%\,\,\,\,\,\,\,\,\,\,{\rm and}\,\,\,\,\,\,\,\,\,\,
% {{\partial n}\over{\partial t}} + n\nabla\cdot \mathbf{v} = -{{\kappa n} \over{w}}\left( \nabla^2 T - {{T}\over{w}} \nabla^2p\right),
\end{equation}
where $\eta$ and $\zeta$ are the shear and bulk viscosity coefficients and $w=e+p$ is the enthalpy density. 
We write the momentum density $\mathbf{M} = \mathbf{g}_l+ \mathbf{g}$, where $\mathbf{\nabla} \times \mathbf{g}_l=0$ and $\mathbf{\nabla} \cdot \mathbf{g}=0$. The shear modes satisfy 
\begin{equation}\label{eq:DiffModes}
 {{\partial }\over{\partial t}}\mathbf{g}   = \nu\nabla^2 \mathbf{g},%\,\,\,\,\,\,\,\,\,\,\,\,\,\,{\rm first~order}
\end{equation}
where $\nu = \eta/w$ is the kinematic viscosity. This is a closed equation. 
Sound modes are curl-free compression waves described by
\begin{equation}\label{eq:LongitudinalMom}
 {{\partial }\over{\partial t}}\mathbf{g}_l + \bm{\nabla} p  = \Gamma_s\bm{\nabla} (\nabla\cdot \mathbf{g}_l)
%\,\,\,\,\,\,{\rm and}\,\,\,\,\,\,\,{{\partial e}\over{\partial t}} + \nabla\cdot \mathbf{g}_l = 0,
\end{equation}
where $\Gamma_s =  (\zeta + 4\eta/3)/w$. We point out that the physics of sound modes is considerably more complex than shear modes, because they also involve pressure fluctuations and heat transfer; see, e.g., \cite{1963AnPhy..24..419K,%forster1995hydrodynamic,
chaikin2000principles}. 

We focus on the damping of transverse velocity fluctuations along the beam direction $z$, which necessarily involves shear modes. More generally, it is useful to understand when shear and other diffusive modes are more important than sound in determining the overall response of the system to perturbations. Equation (\ref{eq:DiffModes})  implies that shear modes of wavenumber $k$ and frequency $\omega$ are damped with $\omega = -i\nu k^2$.  In contrast, sound modes described by (\ref{eq:LongitudinalMom}) propagate at the sound speed $c_s=(\partial p/\partial e)^{1/2}$ with $\omega = \pm c_sk -i\Gamma_s k^2/2\approx \pm c_sk$, where the damping coefficient is $\Gamma_s$ plus thermal conduction contributions. A general perturbation will excite both $\mathbf{g}_l$ and $\mathbf{g}$  at a range of frequencies, and one must consider the combined response.  A low frequency perturbation satisfying
\begin{equation}\label{eq:lowFrequency}
\omega \sim \nu k^2 \ll c_s k, 
%\,\,\,\,\,\,{\rm and}\,\,\,\,\,\,\,{{\partial e}\over{\partial t}} + \nabla\cdot \mathbf{g}_l = 0,
\end{equation}
will predominantly excite shear modes, while perturbations at higher frequencies
\begin{equation}\label{eq:highFrequency}
\omega \sim c_s k \gg \nu k^2 
%\,\,\,\,\,\,{\rm and}\,\,\,\,\,\,\,{{\partial e}\over{\partial t}} + \nabla\cdot \mathbf{g}_l = 0,
\end{equation}
excite sound waves \cite{chaikin2000principles,landau1980statistical}. When hydrodynamics is applicable,  $\nu k\ll c_s$ because macroscopic length scales $\sim k^{-1}$ must greatly exceed the mean free path $\sim \nu$.  We see in secs.\ \ref{sec:constNu} and \ref{sec:betaPlots} that the longitudinal distance scale $k^{-1}$ for rapidity correlations is long and grows with proper time, so that the low frequency regime applies.  

%There are other hydrodynamic modes by which more general fluctuations are propagated and damped. 
 %With $\partial_\mu T^{0\mu} = \partial e/\partial t + \nabla\cdot \mathbf{g}_l = 0$, we see that t

Second order hydrodynamics is widely used in phenomenological studies of nuclear collisions \cite{Romatschke:2009im,Denicol:2014vaa,Betz:2008me}. This formulation is especially important for diffusive phenomena, where it renders the theory causal. In first order diffusion (\ref{eq:DiffModes}), a delta function perturbation instantaneously spreads into a Gaussian, with tails extending to infinity.  %At second order in the mean free path compared to fluid gradients, 
New transport coefficients at second order include relaxation times for shear and bulk stresses, among other terms.  
Linearized forms of the second order equations are discussed in \cite{Romatschke:2009im,Young:2014pka}. To linear order the shear modes satisfy a 
Maxwell-Cattaneo equation
\begin{equation}\label{eq:2DiffModes}
\left(\tau_\pi {{\partial^2}\over{\partial t^2}} + {{\partial }\over{\partial t}}\right)\mathbf{g}   = \nu\nabla^2 \mathbf{g};%\,\,\,\,\,\,\,\,\,\,\,\,\,\,{\rm second~order}
\end{equation}
see eqn.\ (45) in ref.\ \cite{Romatschke:2009im}. Shear modes satisfy $\omega = -i\nu k^2/(1-i\omega\tau_\pi)$, implying that the low-frequency behavior is diffusive, but high frequency pulses can propagate at speeds up to $\sqrt{\nu/\tau_\pi}$.  
%Kinetic theory for massless Boltzmann particles gives $\tau_\pi = 5\eta/w$. 
We stress that this equation only applies for perturbations of a uniform stationary fluid. We will obtain this equation and its generalization to nuclear collisions from the M\"uller-Israel-Stewart equation in sec. \ref{sec:ionCollisions}. 

%Corrections to this Maxwell-Cattaneo 
%form do not contribute to linear order for the problems considered here 

\section{Correlations and Noise}\label{sec:Roadmap}
In a given event, fluctuations in the transverse velocity perturb the shear momentum current of the fluid by an amount $g_i$ in a transverse direction $i=x,y$. We will describe transverse momentum fluctuations in terms of the correlation function
\begin{equation}
r_g^{ij} = \langle g_i(\mathbf{x}_1,t)g_j(\mathbf{x}_2,t)\rangle - \langle
g_i(\mathbf{x}_1,t)\rangle \langle g_j(\mathbf{x}_2,t)\rangle
\label{eq:CcorrF2}
\end{equation}
where the brackets denote an average over an ensemble of possible fluctuations with fixed initial conditions. 
Observables in sec.\  \ref{sec:Observables} are essentially integrals of this function averaged over the physical range of  initial conditions
\cite{Gavin:2006xd}. In local equilibrium, the correlation function (\ref{eq:CcorrF2}) is nonzero due to stochastic hydrodynamic noise.  If we were to omit this noise, $r_g$ would vanish in that limit. We refer to the average in (\ref{eq:CcorrF2}) as the ``noise average'' or the ``thermal average.'' 

In order to calculate correlation functions such as (\ref{eq:CcorrF2}) we must specify: 1) the initial correlations, 2) the hydrodynamic equations and equation of state, 3) the dissipative contributions and transport coefficients, and 4) the hydrodynamic noise.  The first two effects are essential for describing the measured anisotropy of azimuthal flow, and most practitioners also include dissipation. Schematically, the initial correlations are determined by fluctuations in the geometry and number of participants. In each collision event correlated particles are more likely to be found near hot spots produced by these fluctuations. We often associate hot spots with flux tubes produced by the initial nucleon participants, but that association is not essential for this work. 

Hydrodynamic noise is a consequence of the same microscopic scattering processes that produce dissipation. While dissipation tends to dampen the effect of the initial hot spots on pressure and velocity fluctuations, noise opposes this dampening. A number of authors have begun to study theoretical and phenomenological aspects of thermal noise, mostly with the aim of incorporating noise in numerical simulation codes \cite{Calzetta:1997aj,Kapusta:2011gt,Kumar:2013twa,Young:2014pka,Yan:2015lfa,Nagai:2016wyx}.  

In the coming parts of this section, we obtain a partial differential equation for $r_g$ including the effect of noise. We find that $\Delta r_g = r_g - r_{g,\,{\rm le}}$ satisfies a deterministic diffusion equation (\ref{eq:AbdelAzizEquation}) in second order hydrodynamics.  Our result  (\ref{eq:AbdelAzizEquation}) is new and our technique for constructing partial differential equations for correlation functions is unique in the field. In ref.\  \cite{Gavin:2006xd} we used the first order  approximation (\ref{eq:MomDiffusion5}) with only a cursory discussion of the effect of noise. Equation (\ref{eq:AbdelAzizEquation}) refines a more schematic causal diffusion that we used to study net charge and baryon number diffusion in ref.\ \cite{Aziz:2004qu}. We derive (\ref{eq:AbdelAzizEquation}) and discuss the physics at length in part to extend our earlier works to current phenomenological problem. We also hope to develop techniques for applying hydrodynamics to calculate similar correlation functions for other applications.  

To obtain these hydrodynamic equations for correlation functions, we work with stochastic differential equations analytically in a way that is common in probability theory but less familiar in physics.  We will establish these methods heuristically by working from the familiar example of Brownian motion up to diffusion problems more relevant to our system. See ref.\ \cite{van2011stochastic,gardiner2004handbook} for a more detailed treatment.

\subsection{Brownian Motion}

Brownian motion refers to the random zig-zag motion of a heavy particle suspended in a fluid. This motion is described in one dimension by the Langevin equation
%
%\begin{equation}\label{eq:Brownian1}
$m\dot{v} = - m\gamma v  + f,$ 
%\end{equation}
%
where both the friction coefficient $\gamma$ and the stochastic force $f$ are due to collisions with faster-moving fluid particles; we assume non-relativistic motion for this illustration. We write the Langevin equation as a difference equation 
\begin{equation}\label{eq:Brownian1}
v(t+\Delta t)-v(t) \equiv \Delta v =  - \gamma v(t)\Delta t  +  \Delta W, 
\end{equation}
where  $\Delta W$ represents the net change in $v$ due to microscopic collisions in the time interval from $t$ to $t+\Delta t$.  The contribution to $\Delta W$ from each collision is independent and uncorrelated in both direction and magnitude, so that 
\begin{equation}\label{eq:Brownian2}
\langle \Delta W\rangle  = 0\,\,\,\,\,\,\,\,\,\,\,\,  {\rm and} \,\,\,\,\,\,\,\,\,\,\,\, \langle \Delta W^2\rangle  = \Gamma \Delta t, 
\end{equation}
when averaged over the noise, i.e., all possible trajectories of the heavy particle starting with the same velocity $v$ and position $x$. The linear relation $\langle \Delta W^2\rangle  \propto \Delta t$ is typical of random-walk processes and, unopposed by friction, would cause the variance of $v$ to increase in proportion to time \cite{gardiner2004handbook}. We determine the coefficient $\Gamma$ in accord with the fluctuation-dissipation theorem by demanding that fluctuations in equilibrium have the appropriate thermodynamic limit.  

We obtain differential equations for the averages  $\langle v(t)\rangle$ and  $\langle v(t)^2\rangle$ as follows.  
The average of (\ref{eq:Brownian1}) gives $\langle v(t+\Delta t)\rangle - \langle  v(t)\rangle = -\gamma \langle v(t)\rangle \Delta t$, so that
\begin{equation}\label{eq:Brownian2.5}
d{\langle v\rangle}/dt = - \gamma \langle v\rangle. 
\end{equation}
as $\Delta t \rightarrow 0$. In the long time limit, the average $\langle v\rangle$ tends to zero, although each individual particle remains in random motion. The noise term has no effect on the mean. 

In contrast, $\langle v(t)^2\rangle$ is profoundly affected by thermal noise, as is well known. We square (\ref{eq:Brownian1}) to obtain the difference $v(t+\Delta t)^2  -  v(t)^2 = 2  v(t)\Delta v + \Delta v^2$. The average of the first term is $2 \langle v(t)\Delta v\rangle = -2\gamma  \langle v(t)^2\rangle\Delta t$.  We use (\ref{eq:Brownian2}) to average the second term and find $\langle \Delta v^2\rangle = \Gamma \Delta t$ to leading order in $\Delta t$. 
Combining these contributions and taking  $\Delta t \rightarrow 0$, we obtain 
\begin{equation}\label{eq:Brownian3}
d\langle v^2\rangle/dt  =  - 2\gamma \langle v^2\rangle  +  \Gamma.  
\end{equation}
The need to keep $\Delta v^2$ along with $v\Delta v$ in the presence of noise is known in the theory of stochastic differential equations as the It$\hat{\rm{o}}$ product rule. 

In equilibrium the  time derivative in (\ref{eq:Brownian3}) must vanish, so that 
\begin{equation}\label{eq:Brownian4}
\Gamma  =  2\gamma \langle v^2\rangle_{\rm eq}.
\end{equation}
Had we omitted the noise contribution, (\ref{eq:Brownian3}) would give  $\langle v^2\rangle_{\rm eq} = 0$ rather than the equipartition value $\langle v^2\rangle_{\rm eq} = T/m$. We take the equilibrium value to determine $\Gamma  =  2\gamma T/m$. One usually assumes that the particle is always in equilibrium with the fluid, i.e., $\langle v^2\rangle \equiv T/m$, but this need not be the case.

To find the displacement of the Brownian particle, observe that $\Delta x = v(t)\Delta t$ is independent of the noise, so that 
$\Delta (x^2) = 2 x\Delta x = 2 xv\Delta t$. Similarly, $\Delta (xv) = x\Delta v + v\Delta x$, which gives 
\begin{equation}\label{eq:Brownian5}
{{d\langle x^2\rangle}\over{dt}}  =  2\langle xv\rangle   \,\,\,\,\,\, {\rm and}  \,\,\,\,\,\,  {{d\langle xv\rangle}\over{dt}}  =  -\gamma \langle xv\rangle +\langle v^2\rangle. 
\end{equation}
In equilibrium, $\gamma \langle xv\rangle_{\rm eq} = \langle v^2\rangle_{\rm eq} = T/m$ for $T$ the temperature, so that  (\ref{eq:Brownian5}) yields the celebrated random walk result $\langle x^2\rangle =  (2T/\gamma m) t$ for $t \gg \gamma^{-1}$. 

The stopping of fast (but non-relativistic) particles is schematically described by (\ref{eq:Brownian3}). The deviation of the variance $r_v= \langle v^2\rangle -\langle v\rangle^2$ from its equilibrium value measures the degree to which such particles are thermalized by the fluid. Combining the equation of motion for $\langle v(t)\rangle$ with (\ref{eq:Brownian3}), we write
\begin{equation}\label{eq:Brownian6}
d\Delta r_v/dt  =  - 2\gamma \Delta r_v 
\end{equation}
where $\Delta r_v =  r_v - r_{v,\,\rm eq}$ measures the deviation of the variance of  $\langle v^2\rangle$ from its equilibrium value. 

We emphasize two features of (\ref{eq:Brownian6}) common to our key result (\ref{eq:AbdelAzizEquation}).  First, the  relaxation of $\Delta r_v$ to equilibrium is independent of the noise $\Gamma$. Second, the time scale for relaxation of the variance $\Delta r_v$ is $1/2\gamma$ --- half the value for the relaxation of the mean $\langle v\rangle$. This factor is already evident by comparing (\ref{eq:Brownian2.5}) and (\ref{eq:Brownian3}). This factor will be important for our estimate of $\tau_\pi$ in this paper.  

We remark that the propagation of heavy flavor through the quark gluon plasma is often treated with Langevin dynamics \cite{Moore:2004tg,Akamatsu:2015kaa,Song:2015jmn}. Theoretical aspects of relativistic random walks have been addressed using methods similar to ours \cite{Koide:2011yy}. Recent work involves numerical simulations of the relativistic version of (\ref{eq:Brownian1}) with momentum dependent $\gamma$ factors; see, e.g., \cite{He:2013zua}.

\subsection{Particle Diffusion with Noise}
%%%%%%%%%%%%%%%%%%
%%%%%%%%%%%%%%%%%
To generalize this result to hydrodynamics, we start with the first order diffusion equation 
\begin{equation}\label{eq:Diffusion000}
{{\partial n}\over{\partial t}}  =  -\mathbf{\nabla} \cdot \mathbf{J}   \,\,\,\,\,\, {\rm where}  \,\,\,\,\,\,  \mathbf{J} = -D\bm{\nabla} n + \mathbf{j}. 
\end{equation}
The left equation describes number conservation, while the right equation is Fick's law for the current. The new contribution $ \mathbf{j}$ is a stochastic current due to the motion of particles in and out of a fluid cell centered at $\mathbf{x}$. For now we consider only first order hydrodynamics linearized about a stationary background. This is a good starting point because the stochastic diffusion equation is well understood \cite{gardiner2004handbook}. 
 
We write this as a difference equation 
\begin{equation}\label{eq:Diffusion1}
n(t+\Delta t)-n(t) \equiv \Delta n =  D\nabla^2 n(t)\Delta t  +  \Delta W, 
\end{equation}
where $\Delta W$ represents the increment to the density $n$ at the point $\mathbf{x}$ due to  $ \mathbf{j}$ from $t$ to $t+\Delta t$. These increments satisfy $\langle \Delta W(x_1) \Delta W(x_2) \rangle = \Gamma_{12}\Delta t$. The stochastic nature of $ \mathbf{j}$ further implies that $\Delta W(x_i)$ are uncorrelated for $x_1$ and $x_2$ corresponding to different fluid cells. In the hydrodynamic limit where the cell size tends to zero, we therefore expect $\Gamma_{12}$ to be singular at $x_1 = x_2$ and zero otherwise \cite{gardiner2004handbook}. 

As with the previous example, the average of  (\ref{eq:Diffusion1}) satisfies the diffusion equation
\begin{equation}\label{eq:Diffusion2}
{{\partial \langle n\rangle}\over{\partial t}}  =  D\nabla^2 \langle n\rangle. 
\end{equation}
Now consider the correlation function $\langle n(\mathbf{x}_1, t)n(\mathbf{x}_2, t)\rangle \equiv \langle n_1(t)n_2(t)\rangle$. To obtain a differential equation for this correlation function, we write a difference equation for 
%\begin{equation}\label{eq:Diffusion3}
$\Delta \langle n_1n_2\rangle \equiv
\langle n_1(t+\Delta t)n_2(t+\Delta t)\rangle - \langle n_1(t)n_2(t)\rangle.$ 
%\end{equation}
%
We again use It$\hat{\rm{o}}$ product rule:
\begin{equation}\label{eq:Diffusion4}
\Delta \langle n_1n_2\rangle = \langle n_1\Delta n_2\rangle +  \langle n_2\Delta n_1\rangle 
+  \langle \Delta n_1\Delta n_2\rangle, 
\end{equation}
where $ \langle \Delta n_1\Delta n_2\rangle = \Gamma_{12}\Delta t$ is the same order in $\Delta t$ as the other terms owing to its stochastic nature. We combine (\ref{eq:Diffusion1}) and (\ref{eq:Diffusion4})  to obtain
\begin{equation}\label{eq:Diffusion5}
\left[{{\partial }\over{\partial t}}  - D(\nabla_1^2+\nabla_2^2)\right] r_n = \Gamma_{12}, 
\end{equation}
where 
\begin{equation}\label{eq:Diffusion6}
r_n = \langle n_1 n_2\rangle - \langle n_1\rangle\langle n_2\rangle. 
\end{equation}
The local equilibrium correlation function $r_{n,\, {\rm{le}}}$ must be time independent since we have assumed a static background flow.  We must then take $\Gamma_{12} \equiv -D(\nabla_1^2+\nabla_2^2) r_{n,\, {\rm{le}}}$. 

It is useful to eliminate the noise term in (\ref{eq:Diffusion5}) by writing  
\begin{equation}\label{eq:Diffusion5b}
\left[{{\partial }\over{\partial t}}  - D(\nabla_1^2+\nabla_2^2)\right]\Delta r_n = 0, 
\end{equation}
where $\Delta r_n = r_n - r_{n,\, {\rm{le}}}$. Mathematically, this equation is easier to work with than (\ref{eq:Diffusion5}) because $\Gamma_{12}$ is singular at $\mathbf{x_1}=\mathbf{x_2}$; see eq.\  (\ref{eq:NoiseDiff}).  This result is derived more formally in \cite{gardiner2004handbook}. We used a generalization of this equation to study second order net charge correlations in ref.\ \cite{Aziz:2004qu}. 

To determine the local equilibrium $r_{n,\, {\rm{le}}}$, observe that the particle number fluctuations satisfy Poisson statistics when interactions and Bose/Fermi corrections are negligible and the volume under consideration is sufficiently small that the grand canonical ensemble applies. Equilibrium fluctuations then satisfy $\langle N^2\rangle - \langle N\rangle^2 = \langle N\rangle$, which implies that the density correlations  $r_n = \langle \delta n_1 \delta n_2 \rangle$ must equal  $r_{n,\, {\rm{le}}} = \langle n_1\rangle \delta(\mathbf{x}_1 - \mathbf{x}_2)$ in local equilibrium. 

We now obtain the noise term:
\begin{eqnarray}\label{eq:NoiseDiff}
\Gamma_{12} %&=& -D(\nabla_1^2+\nabla_2^2) r_{n,\, {\rm{le}}} \nonumber\\
%&=& 2\mathbf{\nabla}_1\cdot \mathbf{\nabla}_2 D\langle n_1\rangle \delta(\mathbf{x}_1 - \mathbf{x}_2). 
&=& -(\nabla_1^2+\nabla_2^2) D\langle n_1\rangle \delta(\mathbf{x}_1 - \mathbf{x}_2). 
\end{eqnarray}
The presence of noise when $\mathbf{x_1}=\mathbf{x_2}$ due to (\ref{eq:NoiseDiff}) ensures that the particle number within the same fluid cell will fluctuate even in equilibrium. Had we omitted the contribution from noise in (\ref{eq:Diffusion1}), (\ref{eq:Diffusion5}) with $\Gamma_{12}=0$ would predict that $r_n$ would tend to zero instead of $r_{n,\, {\rm{le}}}$ as $t\rightarrow \infty$, in violation of thermodynamics.  

The true utility of (\ref{eq:Diffusion5b}) lies in the fact that $\Delta r_n$ is directly observable by counting particles. The density of distinct pairs is
$\langle n_1 n_2\rangle - \langle n_1\rangle \delta(\mathbf{x}_1 - \mathbf{x}_2)$. In the absence of correlations this density is $\langle n_1\rangle\langle n_2\rangle$. In equilibrium particles at different points are uncorrelated, since $\langle n_1 n_2\rangle =  \langle n_1\rangle\langle n_2\rangle$ except when $\mathbf{x_1}= \mathbf{x_2}$. %By itself, the variance  $\langle n_1 n_2\rangle -  \langle n_1\rangle\langle n_2\rangle$ is singular at  $\mathbf{x_1}=\mathbf{x_2}$ and not directly observable.  
See refs.\ \cite{Pruneau:2002yf} and \cite{Aziz:2004qu} for further discussion of particle correlation measurements. 

We interpret (\ref{eq:Diffusion5b}) as follows.  Suppose that the initial distribution each event is ``clumpy'' with regions of particle surplus and deficit. This inhomogeneity produces spatial correlations, since it is more likely to find particles together near a dense clump. 
%%%%
The spatial size of clumps sets the initial scale of $\Delta r_n$.  As time goes on, (\ref{eq:Diffusion5b}) describes the tendency of diffusion to distribute particles as evenly throughout the volume as possible in the presence of noise.

\subsection{Momentum Diffusion with Noise}
%%%%%%%%%%%%%%%%%%
We start with the first-order momentum diffusion equation as studied in ref.\ \cite{Gavin:2006xd}. Each vector component of the shear contribution to the momentum current satisfies a diffusion equation (\ref{eq:DiffModes}), for which we write difference equations
\begin{equation}\label{eq:MomDiffusion1}
\Delta g^i =  \nu \nabla^2 g^i \Delta t  +  \Delta W^i,
\end{equation}
where $\langle \Delta W(x_1)^i \Delta W(x_2)^j \rangle = \Gamma_{12}^{ij}\Delta t$. 
%We introduce a projection operator $P$ that to ensure that the increment to the shear flow $\Delta W^i$ is divergence free. 
The momentum correlation function, 
\begin{equation}\label{eq:MomDiffusion2}
r_g^{ij} = \langle g_1^i g_2^j\rangle - \langle g_1^i\rangle\langle g_2^j\rangle,
\end{equation}
satisfies the diffusion equation 
\begin{equation}\label{eq:MomDiffusion3}
\left[{{\partial }\over{\partial t}}  - \nu(\nabla_1^2+\nabla_2^2)\right] r_g^{ij} = \Gamma_{12}^{ij}. 
\end{equation}
As before, the noise is fixed to give the correct local equilibrium fluctuations 
\begin{equation}\label{eq:MomDiffusion4}
\Gamma_{12}^{ij} = - \nu(\nabla_1^2+\nabla_2^2)r_{g,\, {\rm{le}}}^{ij},
\end{equation}
where $r_{g,\, {\rm{le}}}^{ij}$ is the equilibrium correlation function.  Note that $\langle g\rangle \equiv 0$ by definition, but we keep this quantity in the calculations for generality. We can then write 
\begin{equation}\label{eq:MomDiffusion5}
\left[{{\partial }\over{\partial t}}  -\nu(\nabla_1^2+\nabla_2^2)\right] \Delta r_g^{ij} = 0. 
\end{equation}
where $\Delta r_g^{ij} = r_g^{ij} - r_{g,\, {\rm{le}}}^{ij}$.   
%%%%%%%%%%%%%%%%%%%%%%

%%%%%%%%%%%%%%%%%%%%%%%%%%
We interpret (\ref{eq:MomDiffusion5}) and its second order extension (\ref{eq:AbdelAzizEquation})  following our discussion of particle diffusion.  
An initially clumpy distribution produces inhomogeneous gradients resulting in anisotropic transverse flow. Viscosity works to reduce the anisotropy, driving $r_g^{ij}$ to $r_{g,\, {\rm{le}}}^{ij}$ the value set by the thermal noise, so that $\Delta r_g^{ij}\rightarrow 0$. 

Generalizations of (\ref{eq:Diffusion5b}) and (\ref{eq:MomDiffusion5}) are phenomenologically useful, so we do not need the explicit forms of $\Gamma_{12}^{ij}$ or $r_{g,\, {\rm{le}}}^{ij}$ to address observations.  That said, we discuss the noise as an aside because of its theoretical interest. Let $f(\mathbf{x}, \mathbf{p},t)$ represent the phase space density in an event, which differs from the thermal average $\langle f(\mathbf{x}, \mathbf{p}, t)\rangle$ by an amount $\delta f = f - \langle f\rangle$.   Poisson statistics requires that $\langle\delta f_1 \delta f_2 \rangle\rightarrow \langle f_1\rangle \delta(\mathbf{x}_1 - \mathbf{x}_2)\delta(\mathbf{p}_1 - \mathbf{p}_2)$ in local equilibrium. The total momentum density excess in an event is $M^i = T^{0i} - \langle T^{0i}\rangle = \int p^i \delta f(x,p) dp$.  The correlation function  $r_M^{ij} = \int p_1^i p_2^j  \langle\delta f_1 \delta f_2 \rangle dp_1dp_2$ has the equilibrium form  $r_{M,\, {\rm{le}}}^{ij} =  A\delta^{ij} \delta(\mathbf{x}_1 - \mathbf{x}_2)$, where $A=wT$. To determine $A= \int (p^i)^2 \langle f\rangle dp$, it suffices to take $v \ll 1$, so that $\langle f\rangle \approx e^{-(E-p\cdot v)/T}$ and $\int p^i \langle f\rangle dp \approx wv^i$. It follows that $w = \int p^i (\partial f/\partial v_i) dp =  \int (p^i)^2 (-\partial f/\partial E) dp= A/T$. 

To obtain the fluctuations of the shear modes $r_g$ from $r_M$, we use (\ref{eq:MomDiffusion2}) to write $r_g^{ij} = P^{i}_l(x_1)P^{j}_m(x_2) r_M^{lm}$, were  $P$ is a linear operator that projects out the divergence-free component of $M$ such that $PM=g$. Equation (\ref{eq:MomDiffusion4}) then yields 
\begin{equation}\label{eq:NoiseMom}
%\Gamma_{12}^{ij} =  -\{\delta^{ij}(\nabla_1^2+\nabla_2^2) -\nabla_1^i\nabla_1^j-\nabla_2^i\nabla_2^j\} \eta T \delta(\mathbf{x}_1 - \mathbf{x}_2).
%\Gamma_{12}^{ij} =  (\nabla_1^i\nabla_1^j-\delta^{ij}\nabla_1^2 + \nabla_2^i\nabla_2^j-\delta^{ij}\nabla_2^2) \eta T \delta(\mathbf{x}_1 - \mathbf{x}_2).
\Gamma_{12}^{ij} =  -(\delta^{ij}\nabla_1^2-\nabla_1^i\nabla_1^j) \eta T \delta(\mathbf{x}_1 - \mathbf{x}_2) + (1\leftrightarrow 2).
%\Gamma_{12}^{ij} =  2(\mathbf{\nabla}_1\cdot \mathbf{\nabla}_2\delta^{ij} -\nabla_1^i\nabla_2^j) \eta T \delta(\mathbf{x}_1 - \mathbf{x}_2).
\end{equation}
Note that the operator $P$ is used in electromagnetism to project out the transverse component of the electric current. 

%where we have made use of the delta functions and the identity $P^2 = P$ for projection operators. %The result is most transparent in Fourier space, where  $\Gamma_{12}^{ij}(k_1, k_2) =  2(2\pi)^3(k_1^2\delta^{ij} -k_1^ik_1^j) \eta T \delta(\mathbf{k}_1 + \mathbf{k}_2)$. 

%
%used $P^{i}_l(x_1)P^{j}_m(x_2)\delta^{lm}\delta(\mathbf{x}_1 - \mathbf{x}_2) = P^{im}(x_1)P^{i}_m(x_1)\delta(\mathbf{x}_1 - \mathbf{x}_2) = P^{ij}(x_1) \delta(\mathbf{x}_1 - \mathbf{x}_2)$, since $P^2 = P$ for projection operators.  

%%%%%%%%%%%
%%
%%                                Chris' corrected formulae
%%
%%%%%%%%%%%

We turn now to the focus of this paper: the diffusion of momentum fluctuations in linearized second order  hydrodynamics. As in Brownian motion, we convert the second order equation (\ref{eq:2DiffModes}) into a first order stochastic system:
\begin{equation}\label{eq:2Diffusion1a}
\Delta h^i =  -\gamma (h^i  -  L g^i) \Delta t  +  \gamma\Delta W^i,
\end{equation}
where $L = \nu\nabla^2$, $\gamma = 1/\tau_\pi$, and 
\begin{equation}\label{eq:2Diffusion1b}
\Delta g^i = h^i\Delta t. 
\end{equation}
Again we keep the quantities $\langle h\rangle$ and $\langle g\rangle$ around for generality, even though they are zero. As in Brownian motion, only the first equation has a stochastic contribution satisfying $\langle \Delta W(x_1)^i \Delta W(x_2)^j \rangle = \Gamma_{12}^{ij}\Delta t$.  For the moment, we hide the vector indices for simplicity.  
We again follow the Brownian motion example, writing
\begin{eqnarray*}\label{eq:2Diffusion2}
\Delta \langle g_1g_2\rangle &=& \langle g_1\Delta g_2\rangle +  \langle g_2\Delta g_1\rangle \nonumber\\
&=& (\langle g_1 h_2\rangle + \langle h_1 g_2\rangle)\Delta t
\end{eqnarray*}
to leading order in $\Delta t$, since (\ref{eq:2Diffusion1b}) is unaffected by noise.  
We define the covariance
\begin{equation}\label{eq:2Diffusion3}
r_{gh} = \langle g_1 h_2\rangle - \langle g_1\rangle\langle h_2\rangle,\,\,\,\, r_{hg} = \langle h_1 g_2 \rangle - \langle h_1\rangle\langle g_2\rangle, 
\end{equation}
and find
\begin{equation}\label{eq:2DiffusionGG}
\frac{\partial}{\partial t} r_g = r_{gh}+r_{hg},
\end{equation}
for $r_g$ defined in  (\ref{eq:MomDiffusion2}). Likewise, we use (\ref{eq:2Diffusion1a}) and (\ref{eq:2Diffusion1b}) to find
\begin{eqnarray*}\label{eq:2Diffusion4}
\Delta \langle g_1h_2\rangle &=& \langle g_1\Delta h_2\rangle +  \langle h_2\Delta g_1\rangle \nonumber\\
&=& (\langle h_1 h_2\rangle -\gamma\langle g_ 1h_2\rangle +\gamma L_2 \langle g_1g_2\rangle )\Delta t,
\end{eqnarray*}
so that
\begin{equation}\label{eq:2DiffusionGH}
\left(\frac{\partial}{\partial t}+\gamma\right) r_{gh} = r_h+ \gamma L_2 r_{g},
\end{equation}
where 
\begin{equation}\label{eq:2Diffusion5}
r_{h} = \langle h_1 h_2\rangle - \langle h_1\rangle\langle h_2\rangle;
\end{equation}
a similar equation for $r_{hg}$ replaces $L_2$ with $L_1$. 
%%
%\begin{equation}\label{eq:2DiffusionHG}
%\left(\frac{\partial}{\partial t}+\gamma\right) r_{hg} = r_h+ \gamma L_1 r_{g}.
%\end{equation}
%%
The sum %and differences of these functions satisfy
of these functions satisfies
\begin{equation}\label{eq:2SumGH}
\left(\frac{\partial}{\partial t}+\gamma\right) (r_{gh}+r_{hg}) = 2r_h+ \gamma (L_1+L_2) r_{g}.
\end{equation}
%%
%and
%%
%\begin{equation}\label{eq:2DiffGH}
%\left(\frac{\partial}{\partial t}+\gamma\right) (r_{gh}-r_{hg}) = \gamma (L_2-L_1) r_{g}.
%\end{equation}
%%
%We see from (\ref{eq:2DiffGH}) that $r_{gh} = r_{hg}$ if correlations are translationally invariant functions of $\mathbf{x}_1 - \mathbf{x}_2$ alone. 

To derive an evolution equation for $r_h$, we must use the It$\hat{\rm{o}}$ product rule:
\begin{equation}\label{eq:2Diffusion6}
\Delta \langle h_1h_2\rangle = \langle h_1\Delta h_2\rangle +  \langle h_2\Delta h_1\rangle 
+  \langle \Delta h_1\Delta h_2\rangle, 
\end{equation}
because $ \langle \Delta h_1\Delta h_2\rangle = \gamma^2\Gamma_{12}\Delta t$ due to the noise contribution to (\ref{eq:2Diffusion1a}). 
We obtain
\begin{equation}\label{eq:2DiffusionHH}
\left(\frac{\partial}{\partial t}+2\gamma\right) r_{h} = \gamma L_1 r_{gh} + \gamma L_2 r_{hg}+ \gamma^2\Gamma_{12}.
\end{equation}
In equilibrium in an infinite system, all the time derivatives vanish and the system is translationally invariant, so that (\ref{eq:2DiffusionGG}) implies $r_{gh,\,{\rm le}} = r_{hg,\,{\rm le}}=0$. Moreover,  (\ref{eq:2DiffusionGH}) and (\ref{eq:2DiffusionHH}) imply
%
%\begin{equation}\label{eq:2DiffusionEquilbrium} 2r_{h,\,{\rm le}} = \gamma\Gamma_{12} \,\,\,\,\,\,\,\,{\rm and}\,\,\,\,\,\,\,\, 2r_{h,\,{\rm le}} = -\gamma(L_1 + L_2)r_{g,\,{\rm le} },
$\gamma\Gamma_{12} = 2r_{h,\,{\rm le}}  = -\gamma(L_1 + L_2)r_{g,\,{\rm le} },$
%\end{equation}
%
so that  $\Gamma_{12}$ is given by (\ref{eq:NoiseMom}). 

We again introduce $\Delta r_{g} = r_{g}-r_{g, \,{\rm le}}$ and $\Delta r_{h} = r_{h}-r_{h,\,{\rm le}}$ in (\ref{eq:2DiffusionGG}), (\ref{eq:2SumGH}), and (\ref{eq:2DiffusionHH}) to eliminate the $\Gamma_{12}$ contributions. Next we use (\ref{eq:2DiffusionGG}) to eliminate $r_{gh}+r_{hg}$ in (\ref{eq:2SumGH}) and find
\begin{equation}\label{eq:2DiffusionTemp1}
\left(\frac{\partial}{\partial t}+\gamma\right) \frac{\partial}{\partial t}\Delta r_{g} = 2\Delta r_h + \gamma (L_1 + L_2)\Delta r_{g}.
\end{equation}
%
%where we again eliminate the sum using (\ref{eq:2DiffusionGG}). 
We then write
\begin{eqnarray}\label{eq:2DiffusionTemp2}
\left(\frac{\partial}{\partial t} + 2\gamma\right) \Delta r_{h} &=& \tfrac{1}{2}\gamma(L_1 + L_2)(r_{gh}+r_{hg})%\frac{\partial}{\partial t}\Delta r_{g} 
\nonumber\\
&+&\tfrac{1}{2} \gamma(L_1 - L_2)(r_{gh}-r_{hg}).
\end{eqnarray}
Taking $\partial/\partial t + 2\gamma$ on (\ref{eq:2DiffusionTemp1})  and using (\ref{eq:2DiffusionGG}), we find
\begin{eqnarray}\label{eq:2DiffusionTemp3}
\left(\frac{\partial}{\partial t}+\gamma\right)\left[\frac{\partial^2}{\partial t^2} + 2\gamma\frac{\partial}{\partial t} -2\gamma(L_1+L_2)\right] \Delta r_{g}  \nonumber\\
= \gamma(L_1-L_2)(r_{gh}-r_{hg}).
\end{eqnarray}
The difference $r_{gh}-r_{hg}$ on the right side of (\ref{eq:2DiffusionTemp3}) satisfies 
\begin{equation}\label{eq:2DiffGH}
\left(\frac{\partial}{\partial t}+\gamma\right) (r_{gh}-r_{hg}) = \gamma (L_2-L_1) \Delta r_{g},
\end{equation}
where we use (\ref{eq:2DiffusionGH}) and take $L_1r_{g, \,{\rm le}}= L_2r_{g, \,{\rm le}}$, since the local equilibrium distribution $r_{g, \,{\rm le}}$ is translationally invariant. 

We can appreciably simplify (\ref{eq:2DiffusionTemp3}) if  $r_{gh} \approx r_{hg}$, as follows when the right side of  (\ref{eq:2DiffGH}) is negligible. To see when this is the case, write $L_{1,2}$ in terms of the relative and average coordinates,
$\mathbf{x}_r= \mathbf{x}_1 - \mathbf{x}_2$ and $\mathbf{x}_a= (\mathbf{x}_1 + \mathbf{x}_2)/2$, respectively. Then 
\begin{equation}\label{eq:chrisz}
L_1 + L_2 = 2\nu\nabla_r^2 + \frac{\nu}{2}\nabla_a^2 \,\,\,\,\,{\rm and }\,\,\,\,L_1-L_2 = 2\nu{\nabla}_a\cdot{\nabla}_r.
\end{equation}
The right side of (\ref{eq:2DiffGH}) is zero if correlations are translationally invariant, so that they only depend on $\mathbf{x}_r$. If correlations are symmetric in $\mathbf{x}_a$ and slowly varying compared to $\mathbf{x}_r$ then (\ref{eq:2DiffGH}) is negligible near $\mathbf{x}_a = 0$. This holds for the situations we consider in this paper. Finally, if we average the correlation functions over the full range of $\mathbf{x}_a$ then the right side of (\ref{eq:chrisz}) contributes a surface term that must vanish.  

We will solve the approximate evolution equation 
\begin{equation}\label{eq:AbdelAzizEquation}
\left[\frac{\tau_\pi}{2}\frac{\partial^2}{\partial t^2} + \frac{\partial}{\partial t} -\nu(\nabla_1^2+\nabla_2^2)\right] \Delta r_{g}  =0,
\end{equation}
in which we restore the original notation.  This equation is a solution of (\ref{eq:2DiffusionTemp3}) for $r_{gh} = r_{hg}$  in an infinite system. Equation (\ref{eq:AbdelAzizEquation}) is hyperbolic, so that an initial pulse will propagate as a wave, as noted in sec.\ \ref{sec:Modes}. It is also a relaxation equation
\begin{equation}\label{eq:2VisDiffRelax}
 \frac{\partial}{\partial t}\Psi = -\frac{2}{\tau_\pi}\left[ \Psi -  \nu(\nabla_1^2+\nabla_2^2)\Delta r_g \right],
  %\left(\mathbf{x}_1,\mathbf{x}_2,\tau\right)
\end{equation}
where $\Psi = \partial (\Delta r_g) /\partial t$. For $t \gg \tau_\pi/2$, the Navier-Stokes first order diffusion equation (\ref{eq:MomDiffusion5}) holds. The halving of the relaxation time compared to the mean behavior described by (\ref{eq:2DiffModes}) is precisely the same behavior we saw in Brownian motion; see eq. (\ref{eq:Brownian6}).

We comment that there are two cases where we may need to solve the coupled equations (\ref{eq:2DiffusionTemp1}), (\ref{eq:2DiffusionTemp2}) and (\ref{eq:2DiffGH}) rather than (\ref{eq:AbdelAzizEquation}). In considering the rapidity dependence in an asymmetric $pA$ collision, there may be an interesting asymmetric $\mathbf{x}_a$ dependence.  However, we can also remove this dependence by averaging over $\mathbf{x}_a$. Alternatively, if the coefficients are strongly time or position dependent, then the derivation of (\ref{eq:AbdelAzizEquation}) will not hold.  

%
%%%%%%%%%%%%%%%  SECTION:              ION COLLISIONS
%
\section{Ion Collisions}\label{sec:ionCollisions}
In this section we apply our formulation to the diffusion of transverse momentum fluctuations through the expanding fluid produced in a nuclear collision. 
Such fluctuations are driven by the shear modes we have been discussing. 
We begin by summarizing the relevant relativistic hydrodynamic equations. For simplicity, we omit contributions from bulk viscosity and thermal conduction as they do not affect the shear modes. To set the pattern for the rest of this section, we derive (\ref{eq:2DiffModes}) describing shear perturbations of a static fluid.  We then develop a linearized hydrodynamic description of  fluctuations about a fluid with an underlying Bjorken flow \cite{Young:2014pka}. After obtaining the familiar equations describing the thermal-averaged underlying flow, we generalize (\ref{eq:AbdelAzizEquation}) for the fluctuations of that flow.

Recall that in relativistic hydrodynamics the state of the fluid is characterized by the local energy density $e$, pressure $p$, and four velocity $u^{\mu} = \gamma(1, \mathbf{v})$ for $\gamma = (1-v^2)^{-1/2}$ at each space-time point. The equations of motion of the fluid are determined by energy-momentum conservation $\partial_\mu T^{\mu\nu} = 0$. The stress-energy tensor for an ideal dissipation-free fluid is $T_{\rm id}^{\mu\nu} = (e+p)u^\mu u^\nu - p g^{\mu\nu}$. More generally, $T^{\mu\nu} = T_{\rm id}^{\mu\nu} + \Pi^{\mu\nu}$,  where $\Pi^{\mu\nu}$ describes the deviation from ideal behavior due to viscosity and other dissipative processes. Such processes arise when the mean free path of individual particles approach the space-time scales over which $e$, $p$, and $u^\mu$ vary. We therefore express $\Pi^{\mu\nu}$ using the co-moving time derivative and gradient 
\begin{equation}\label{eq:Derivatives}
D = u^{\mu}\partial_{\mu}  \,\,\,\,\,\,\,\,\,\,{\rm and}\,\,\,\,\,\,\,\,\,\, \nabla_{\mu} = \partial_\mu - u_\mu u^\nu \partial_\nu,
\end{equation}
for the metric $g^{\mu\nu}={\rm diag}(1,-1,-1,-1)$. 
In the local rest frame where $u_\mu = (1,0,0,0)$, these quantities are the time derivative and three-gradient. 

At first order in the mean free path,  $\Pi^{\mu\nu}$ is given by the shear stress 
\begin{equation}\label{eq:NavStokes1}
S^{\mu\nu} = \eta\left(\nabla^\mu u^\nu + \nabla^\nu u^\mu -\frac{2}{3}\Delta^{\mu\nu} \nabla_\alpha u^\alpha \right),
%\,\,\,\,\,\,\,\,\,\,{\rm and}\,\,\,\,\,\,\,\,\,\,
% {{\partial n}\over{\partial t}} + n\nabla\cdot \mathbf{v} = -{{\kappa n} \over{w}}\left( \nabla^2 T - {{T}\over{w}} \nabla^2p\right),
\end{equation}
where $\eta$ is the shear viscosity coefficient  and we use the Landau-Lifshitz definition of the four velocity. The projector $\Delta^{\mu\nu} = g^{\mu\nu}-u^{\mu}u^{\nu}$ satisfies $\Delta^{\mu\nu}u_\nu = 0$.  Requiring $\partial_\mu (T_{\rm id}^{\mu i} + S^{\mu i}) = 0$ for each spatial direction $i$ yields the Navier-Stokes equation.

In second order hydrodynamics, one writes relaxation equations for the shear stress, bulk stress and heat current \cite{Betz:2008me,Denicol:2010tr,Denicol:2014vaa}.  %Bulk viscosity and thermal conduction do not contribute to the shear flow.   We omit these processes to write 
We keep only the shear contribution, which satisfies% and write
\begin{equation}\label{eq:ISequation}
\Delta^{\mu}_{\alpha} \Delta^{\nu}_{\beta} D \Pi^{\alpha\beta} = -\frac{1}{\tau_\pi}(\Pi^{\mu\nu}  -S^{\mu\nu} ) - \kappa \nabla_\alpha u^\alpha \Pi^{\mu\nu},
\end{equation}
where $\tau_\pi$ is the shear relaxation time and $\kappa$ is given by (\ref{eq:kappa}). This M\"uller-Israel-Stewart equation has seen wide use, see e.g., \cite{Young:2014pka}. %We remark that additional transport coefficients can contribute to the right side of (\ref{eq:ISequation}), but they do not qualitatively alter our linearized equations.  

To illustrate how shear modes evolve in second order hydrodynamics, consider the fluctuations of a fluid that is for the most part at rest. We take the momentum current  $M^i \approx (e+p)v^i$ to be small, so that momentum conservation $\partial_\mu T^{\mu i} = 0$ implies
\begin{equation}\label{eq:bj00}
\frac {\partial}{\partial t}M^i+ \partial^i p= -{\partial}_{\mu} \Pi^{\mu i},
\end{equation}
to linear order in $\mathbf{v}$ and $\mathbf{M}$. Similarly, (\ref{eq:ISequation}) gives
\begin{equation}\label{eq:deltaISequation00}
\frac {\partial}{\partial t}  \Pi^{\mu i} = -\frac{1}{\tau_\pi}( \Pi^{\mu i}  - S^{\mu i} )
\end{equation}
to this order. As in sec.\ \ref{sec:Modes} we write $M^i = g_l^i + g^i$ where $g^i$ is the divergence-free shear current. The curl-free current $g_l^i$ can be expressed as the gradient of a potential and, consequently, receives contributions only from terms proportional to $\partial^i$ in (\ref{eq:bj00}) and (\ref{eq:deltaISequation00}). 
Discarding these terms, we write  
\begin{equation}%\label{eq:bjtemp}
\frac {\partial}{\partial t}g^i= -{\partial}_{\mu} \Pi_T^{\mu i} \,\,\,\,\,\,\,\,\,\,{\rm and}\,\,\,\,\,\,\,\,\,\, \frac {\partial}{\partial t}  \Pi_T^{\mu i} = -\frac{1}{\tau_\pi}( \Pi_T^{\mu i}  -S_T^{\mu i} ),\nonumber
\end{equation}
where $\Pi_T^{\mu i}$ and  $S_T^{\mu i}$ include only the shear contributions. Taking the time derivative of the left equation and the gradient of the right, we eliminate $\Pi_T^{\mu i}$ to find
\begin{equation}
\left(\tau_\pi {{\partial^2}\over{\partial t^2}} + {{\partial }\over{\partial t}}\right)g^i   = -\partial_\mu S_T^{\mu i}. \nonumber
\end{equation}
Linearizing (\ref{eq:NavStokes1}) for perturbations of a fluid at rest, we find $\partial_\mu S_T^{\mu i}= - \nu \nabla^2 g^i$, which gives (\ref{eq:2DiffModes}).
This equation holds only for fluctuations of a quiescent fluid, and has been derived by different methods elsewhere \cite{Romatschke:2009im}.  

In nuclear collisions, hydrodynamic noise produces small variations of the momentum current $M^i = T^{0i}-\langle T^{0i}\rangle$ in each event compared to the average over this noise. We assume the average flow velocity has the Bjorken form, $u^\mu=(t/\tau,0,0,z/\tau)$, where $\tau =(t^2 -z^2)^{1/2}$ and $\eta = (1/2) \log ((t+z)/(t-z))$. The average entropy density $s$ then evolves as a function of proper time following the set of evolution equations first derived in ref.\ \cite{Muronga:2001zk} (and the erratum). We take $u_{\nu}\partial_\mu (T_{\rm id}^{\mu\nu} + \Pi^{\mu\nu}) =0$ and use  $u_{\nu}\partial_\mu T_{\rm id}^{\mu\nu} = De+(e+p)\nabla_\mu u^\mu$. Bjorken flow implies $\nabla_\mu u^\mu = 1/\tau$ and $u^\mu\partial_\mu = \partial/\partial\tau$, while $de = Tds$ and $e+p = Ts$ at zero net baryon density. We find    
\begin{equation}\label{eq:entdiff}
 \frac{ds}{d\tau}+\frac{s}{\tau} = \frac{\Phi}{T\tau},
\end{equation}
where $\Phi = \Pi^{zz}$. The entropy density evolves due to longitudinal expansion and viscous heating. Causality delays the heating following the relaxation equation (\ref{eq:ISequation}), which implies
\begin{equation}\label{eq:entheat}
 \frac{d\Phi}{d\tau} = -\frac{1}{\tau_\pi}\left( \Phi - \frac{4\eta}{3\tau} \right)
 -\frac{\kappa}{\tau} \Phi.
 % -\frac{\Phi}{2}\left[ \frac{1}{\tau} + 
 %\frac{\eta T}{\tau_\pi}\frac{d}{d\tau}\left( \frac{\tau_\pi}{\eta T}\right)\right]
\end{equation}
The coefficient $\kappa$ is 
\begin{equation}\label{eq:kappa}
\kappa = \frac{1}{2}\left\{ 1 + \frac {d\ln(\tau_\pi/\eta T)}{d\ln\tau}\right\}. 
%\approx \frac{1}{2}\left\{ 1 - c_s^2 \frac {d\ln(\tau_\pi/\eta T)}{d\ln T}   \right\},
\end{equation}
%
%We can write $\kappa \approx (1/2)\{ 1 - c_s^2 d\ln(\tau_\pi/\eta T)/d\ln T\}$, using (\ref{eq:entdiff}) to leading order in the transport coefficients and recognizing $sdT/Tds = dp/de = c_s^2$ as the square of the sound speed. To relate $\eta$, $T$, and $s$ in (\ref{eq:entdiff}) and (\ref{eq:entheat}) we need an equation of state and a relationship between shear viscosity $\eta$ with $T$ or $s$.  
For a conformal liquid in which the only scale is $T$, $\tau_\pi\sim 1/T$ and $\eta\sim s\sim T^3$ give $\kappa = 4/3$.

We next study fluctuations relative to this mean flow, focusing on the longitudinal variation of transverse flow fluctuations. To generalize (\ref{eq:bj00}), we compute $\partial_\mu (\delta T_{\rm id}^{\mu i} + \delta \Pi^{\mu i}) = 0$ including the underlying expansion in the first term to obtain
\begin{equation}\label{eq:bj0}
\left(\frac {\partial}{\partial \tau}+\frac{1}{\tau}\right)M^i + \partial^i p=
%\frac {1}{\tau} \frac {\partial}{\partial \tau} (\tau g^i) =
-{\partial}_{\mu} \delta \Pi^{\mu i},
\end{equation}
where we take $M^i = \delta T_{\rm id}^{0 i}$ for $i = x, y$ the Cartesian transverse coordinates. % as in eq.\ (\ref{eq:NavStokes2}). 
Linearizing the relaxation equation (\ref{eq:ISequation}) following  ref.\ \cite{Young:2014pka} gives,
\begin{equation}\label{eq:deltaISequation}
D \delta \Pi^{\mu i} = -\frac{1}{\tau_\pi}(\delta \Pi^{\mu i}  - \delta S^{\mu i} ) - \frac{\kappa}{\tau} \delta \Pi^{\mu i}, 
\end{equation}
where we eliminate many of the terms by using the Bjorken-flow identities $\tilde\nabla^\mu \tau = (t/\tau, 0, 0, z/\tau) = u^{\mu}$ and $\partial^\mu u^\nu = \tilde\Delta^{\mu\nu}/\tau$, where the velocity projector $\tilde\Delta^{\mu\nu}$ only has non-zero $t$ and $z$ components \cite{Bjorken:1982qr}. Equation (\ref{eq:deltaISequation}) includes a $\kappa$ term absent in (\ref{eq:deltaISequation00}) because $\nabla_\alpha u^\alpha = 1/\tau$.

The shear contribution to $M^i$ must be divergence-free, so that (\ref{eq:bj0}) implies 
\begin{equation}\label{eq:bjtemp}
\left(\frac {\partial}{\partial \tau}+\frac{1}{\tau}\right)g^i =
-{\partial}_{\mu} \delta \Pi_T^{\mu i}.
\end{equation}
The divergence-free contribution $\delta \Pi_T^{\mu i}$ satisfies (\ref{eq:deltaISequation}) with $\delta S^{\mu i}$ replaced by $\delta S_T^{\mu i}$. Linearizing (\ref{eq:NavStokes1}) for Bjorken flow gives $\delta S_T^{\mu i}=\eta {\tilde\nabla}^\mu\delta u^i =\nu  {\tilde\nabla}^\mu g^i$, where $\tilde\nabla_{\mu}$ refers to the gradient co-moving with the Bjorken flow. %We find ${\tilde\nabla}^{2}=\tau^{-2}\partial^2 /\partial\eta^2$. 

As with the static background eqs.\ (\ref{eq:bj00}) and (\ref{eq:deltaISequation00}), we seek to obtain an equation for $g^i$ by using (\ref{eq:deltaISequation}) and (\ref{eq:bjtemp}) to eliminate $\delta \Pi_T^{\mu i}$.  Observe that $u_{\mu}\delta \Pi_T^{\mu i}= 0$ for Bjorken flow, while (\ref{eq:deltaISequation}) further implies that $u_{\mu}D\delta \Pi_T^{\mu i} = 0$. Equation (\ref{eq:bjtemp}) then reduces to 
\begin{equation}\label{eq:bj1}
\left(\frac {\partial}{\partial \tau}+\frac{1}{\tau}\right)g^i =
\frac {1}{\tau} \frac {\partial}{\partial \tau} ( g^i\tau) =
-{\tilde\nabla}_{\mu} \delta \Pi^{\mu i},
\end{equation}
where we have used the identity ${\partial}_{\mu} = u^\mu D + {\nabla}_\mu$. Next, we take the gradient $\tilde\nabla_{\mu} = \tilde\Delta^{\mu\nu}\partial_{\nu}$ of (\ref{eq:deltaISequation}).  Using 
\begin{eqnarray*}%\label{eq:leftside}
\tilde \nabla_{\mu} (u^{\nu} \partial _{\nu}\delta \Pi^{\mu i})  &=&
(\tilde\nabla_{\mu} u^{\nu})(\partial_{\nu} \delta \Pi^{\mu i}) + u^{\nu}
\partial_{\nu}(\tilde\nabla_{\mu} \delta \Pi^{\mu i})\nonumber\\
   &=&\frac{1}{\tau} \tilde\nabla_{\mu} \delta \Pi^{\mu i} + \frac {\partial}{\partial
\tau}(\tilde\nabla_{\mu}\delta \Pi^{\mu i}),
\end{eqnarray*}
%
%while on the Navier-Stokes term on the right side becomes $\tilde\nabla_{\mu}\delta S^{\mu i} = \nu {\tilde\nabla}^2 g^i$, where ${\tilde\nabla}^{2}=\tau^{-2}\partial^2 /\partial\eta^2$. We obtain
we find
\begin{equation}\label{eq:bj3}
\left(\frac {\partial}{\partial \tau} +\frac{1}{\tau_\pi} + \frac{\kappa}{\tau}\right)
(\tau{\tilde\nabla}_{\mu}\delta \Pi^{\mu i}) = \frac{\nu}{\tau_\pi}({\tilde\nabla}^2g^i\tau).
\end{equation}
Together, (\ref{eq:bj1}) and (\ref{eq:bj3}) describe the longitudinal diffusion of transverse flow fluctuations of Bjorken average flow. 

%%%%%%%%%%%%%%%%%%%%%%%%%%
%%%%%%%%%%%%%%%%%%%%%%%%%%
%%%%%%%%%%%%%%%%%%%%%%%%%%

To obtain an equation analogous to (\ref{eq:2DiffModes}) for the expanding system, observe that the rapidity density of total momentum $G^i\equiv \int  g^i \tau dx_{\bot}$, where the integral is over the transverse area of the two colliding nuclei. If one identifies spatial
rapidity $\eta$ with the momentum-space rapidity of particles, then $G^i$ is observable. We combine  (\ref{eq:bj1}) and (\ref{eq:bj3}) to find that this rapidity density satisfies
\begin{equation}\label{eq:bj4}
\tau_\pi\frac{\partial ^{2} G^i}{\partial \tau^{2}} +
\left(1+ \frac{\kappa\tau_\pi}{\tau} \right)\frac{\partial G^i}{\partial\tau} =\frac{\nu}{\tau^{2}} \frac
{\partial ^{2} G^i}{\partial \eta^{2}}.
\end{equation}
In the absence of diffusion, the rapidity density $G^i$ remains constant although the spatial density $g^i$ changes due to the underlying Bjorken expansion. Diffusion tends to broaden the rapidity dependence of $G^i$. This equation is modified by noise as in the previous sections. To define the Langevin force in the evolving system, observe that noise is due to the microscopic motion within a fluid cell. Hydrodynamics applies when each cell is effectively point like. Therefore, we define the noise in the local rest frame of the cell using (\ref{eq:NoiseMom}) and transform to rapidity coordinates. Nevertheless, care would be needed to treat a cell of finite size because, e.g., it takes time for fluctuations to propagate across a cell. Even for simple Brownian motion, it would take time for a heavy particle of finite size to respond to random collisions. This is an interesting problem for future research \cite{Dunkel:2009tla}. 

We finally obtain the second order viscous diffusion equation for transverse momentum correlations in rapidity
\begin{equation}\label{eq:2VisDiff}
 \left[ \frac{\tau_\pi^*}{2}\frac{\partial^2}{\partial \tau^2}
%  +\frac{\partial}{\partial \tau}\right)\Delta r_G^{ij} = 
  +\frac{\partial}{\partial \tau}
  -\frac{\nu^*}{\tau^2}\left( 2\frac{\partial^2}{\partial \eta_r^2} + \frac{1}{2}\frac{\partial^2}{\partial \eta_a^2} \right) \right]
  \Delta r_G^{ij} = 0,
  %\left(\mathbf{x}_1,\mathbf{x}_2,\tau\right)
\end{equation}
where 
\begin{equation}\label{eq:RGdef}
r_G^{ij} = \langle G_1^i G_2^j\rangle - \langle G_1^i\rangle\langle G_2^j\rangle
\end{equation}
and $\Delta r_G^{ij}$ is the difference of $r_G^{ij}$ from its equilibrium value $r^{ij}_{G,\,\rm le}$. We stress that $\Delta r_G^{ij}$ is unaffected by the noise. For later convenience we use the relative rapidity $\eta_r \equiv \eta_1 -\eta_2$ and average rapidity $\eta_a=(\eta_1 + \eta_2)/2$ in (\ref{eq:RGdef}). In deriving (\ref{eq:2VisDiff}) we start with (\ref{eq:bj4}) and absorb the effect of $\kappa$  by defining $\tau_\pi^*=\tau_\pi/(1+\kappa \tau_\pi/\tau)$ and $\eta^*=\eta/(1+\kappa \tau_\pi/\tau)$. We then follow the derivation of (\ref{eq:AbdelAzizEquation}), taking the coefficients to vary slowly with time. To be sure,
the coefficients also depend on time because $\tau_\pi$ and $\eta$ vary with the mean temperature obtained from (\ref{eq:entdiff}) and (\ref{eq:entheat}). To strictly account for the time dependence, one may solve a family of coupled equations (\ref{eq:2DiffusionTemp1}), (\ref{eq:2DiffusionTemp2}) and (\ref{eq:2DiffGH}). We feel that (\ref{eq:2VisDiff}) is adequate for our exploratory study.

As noted earlier, (\ref{eq:2VisDiff}) is a hyperbolic wave equation. Because it can also be written in the form (\ref{eq:2VisDiffRelax}), it relaxes to a diffusion equation  
\begin{equation}\label{eq:2VisDiffRelax2}
 \frac{\partial}{\partial \tau} \Delta r_G^{ij}\approx \frac{\nu^*}{\tau^2}\left( 2\frac{\partial^2}{\partial \eta_r^2} + \frac{1}{2}\frac{\partial^2}{\partial \eta_a^2} \right) \Delta r_G^{ij},
  %\left(\mathbf{x}_1,\mathbf{x}_2,\tau\right)
\end{equation}
for $\tau \gg \tau_\pi/2$, except near the wave font, where the second time derivative is always important. The temperature and time variation of coefficients as well as the explicit $\tau$ dependence of (\ref{eq:2VisDiffRelax2}) affect the relaxation rate. We see that the relaxation equation (\ref{eq:entheat}) has a similar form for $\Phi$, implying that the stress energy tensor tends to the Navier-Stokes form for $\tau \gg \tau_\pi$, with a similar caveat about the $\tau$ dependance.   This behavior will have important observable consequences in sec.\ \ref{sec:constNu}.

%equation of state is Hirano-Guylassy (first order phase transition) \cite{Hirano:2005wx} or  
%To relate shear viscosity $\eta$ to $T$ and $s$ we follow \cite{Hirano:2005wx} again in the case of the first order transition. When using the Lattice m we follow \cite{Niemi:2012ry, Niemi:2011ix}. 

\section{Observables}\label{sec:Observables}
The diffusion of transverse momentum correlations can be observed by measuring   
the covariance
\begin{equation}\label{Cdef}
   {\cal C}_{ij} =  \langle N\rangle^{-2}\langle \sum_{a\neq b} p_{i,a}p_{j,b}\rangle 
   -\langle p_i\rangle \langle  p_j\rangle,
\end{equation}
where $i$ and $j$ label the vector components of the momentum,  $a$ and $b$ label particles from each event, and the brackets
here represent the event average. The average momentum is $\langle p_i\rangle \equiv \langle \sum_a p_{i,a}\rangle/\langle
N\rangle$. In the absence of correlations ${\cal C}_{ij} = 0$, as is the case for local equilibrium in an infinite system.  

In this section we must distinguish averages over events from the noise averages used exclusively in the previous sections. Here, we denote the event average of $X$ by $\langle X\rangle$ and the noise average as $\langle X\rangle_n$.  Mathematically, event averages of a noise-averaged quantity $\langle \langle X\rangle_n\rangle$ amount to averages over the initial conditions for $\langle X\rangle_n$.  

The covariance ${\cal C}_{ij}$ for Cartesian transverse components $i = x, y$ measures the fluctuations of conserved quantities: the components of total transverse momentum \cite{Pratt:2010zn}.  Such fluctuations are highly constrained, as we see by considering an ideal measurement that detects all particles with perfect efficiency.  There are no fluctuations in this limit, because each component of the total momentum $P_i\equiv\sum_{a} p_{i,a} $ vanishes in every event.   The unrestricted sum over pairs $\sum_{a, b} p_{i,a}p_{j,b} = P_iP_j$ also vanishes, so that $\sum_{a\neq b} p_{i,a}p_{j,b} =  -  \sum_{a} p_{i,a}p_{j,a} $. It follows from (\ref{Cdef}) that 
\begin{equation}\label{Cfull}
   {\cal C}_{ij} \rightarrow  -\frac{\langle p_i^2\rangle}{\langle N\rangle} \delta_{ij} %,\,\,\,\,\,\,\,\,\,\,\,\,\,\,{\rm all~particles}
\end{equation}
%
%${\cal C}_{ij}\rightarrow  -(\langle p_i^2\rangle/\langle N\rangle) \delta_{ij}$ 
for all particles in the full $\eta_r$ range. We point out that fluctuations of conserved quantities have been studied in many contexts; see e.g., \cite{Pratt:2015jsa}. 

Measurements of ${\cal C}_{ij}$ in a finite rapidity interval differ from (\ref{Cfull}) because conserving particles fall outside the interval. Our interest lies in finding the mechanisms that transport them outside that interval. Transverse momentum is distributed over a large rapidity range early in the collision by Glasma fields together with jet, minijet and string fragmentation processes. Subsequent evolution is more local, involving particle scattering and, ultimately, diffusion.  Measurement of ${\cal C}_{ij}$ probes these rapidity scales. Furthermore, the evolution of azimuthal anisotropy can also be studied using $\gamma^\prime \equiv({\cal C}_{yy} - {\cal C}_{xx})/({\cal C}_{yy} + {\cal C}_{xx})$, as proposed in ref.\ \cite{Pratt:2010zn}. 
 
The covariance in a rapidity interval is related to the spatial correlation function (\ref{eq:RGdef}) by
\begin{equation}\label{CcorrF}
   {\cal C}_{ij} = \langle N\rangle^{-2}\int  \langle \Delta r_G^{ij}(\eta_r, \eta_a)\rangle 
   d\eta_r d\eta_a.
\end{equation}
The brackets in $\langle \Delta r_G^{ij}\rangle$ remind us that this quantity is first averaged over the noise as in (\ref{eq:CcorrF2}), and then over the initial conditions, corresponding to a true event average.   

The result (\ref{CcorrF}) was first obtained in ref.\ \cite{Gavin:2006xd}. Here, we expand the arguments to clarify the approximations.   
Consider $\delta f(\mathbf{x}, \mathbf{p},t)$, the difference of the phase space distribution in an event from the noise-averaged $\langle f\rangle_n$. The contribution of fluctuations to the transverse momentum current is
%
%\begin{equation}\label{momentum_density}
 $M_i(\mathbf{x}) = \int \delta f(\mathbf{x},\mathbf{p})p_i dp$.
%\end{equation}
%
Fluctuations contribute to the unrestricted sum $\sum_{a,b} p_{i,a}p_{j,b} = \langle \int p_{i1}p_{j2} dn_1dn_2\rangle_n =P_iP_j+\int \langle M_i(\mathbf{x}_1)M_j(\mathbf{x}_2)\rangle_n dx_1 dx_2$. 
%
%
%
%, but not to the single-particle average 
%$\langle N\rangle\langle p_i\rangle = \int  p_i\langle f\rangle dpdx$. 
Averaging this quantity over events yields
$\langle \sum_{a,b} p_{i,a}p_{j,b} \rangle =\langle P_iP_j\rangle + \int \langle\langle M_i(\mathbf{x}_1)M_j(\mathbf{x}_2)\rangle_n\rangle dx_1 dx_2$. 
We assume that freeze out occurs at a constant proper time within the collision volume, so that limiting the space integrals to a spatial rapidity interval gives $\int \langle\langle M_i(\mathbf{x}_1) M_j(\mathbf{x}_2)\rangle_n\rangle dx_1 dx_2 = \int \langle\langle {\cal M}_i(\eta_1) {\cal M}_j(\eta_2)\rangle_n\rangle d\eta_r d\eta_a$, where ${\cal{M}}_i = \int M_i \tau dx_\perp$ is the rapidity density of transverse momentum.

Our physics arguments suggest that shear modes $r^{ij}_G$ in (\ref{eq:RGdef}) drive the correlations of ${\cal{M}}$; we prove this shortly. 
For now, we identify  $\int \langle\langle {\cal M}_i(\eta_1) {\cal M}_j(\eta_2)\rangle_n\rangle d\eta_r d\eta_a = \int \langle r^{ij}_G \rangle d\eta_r d\eta_a$, so that
\begin{equation}\label{unres0}
 \int \langle r^{ij}_G\rangle d\eta_r d\eta_a 
 =\langle \sum_{{\rm all}\, a,b} p_{i,a}p_{j,b}\rangle - \langle P_iP_j\rangle.
 %\langle N\rangle \langle p_i\rangle \langle  p_j\rangle
\end{equation}
We use (\ref{Cdef}) to write the unrestricted sum as
\begin{equation}\label{unres1}
\langle \sum_{{\rm all}\, a,b} p_{i,a}p_{j,b}\rangle
    = \langle N\rangle^2{\cal C}_{ij}+\langle\sum_{a} p_{i,a}p_{j,a}\rangle
    +\langle P_i\rangle\langle P_j\rangle,
\end{equation}
where we added and subtracted $\langle P_i\rangle\langle P_j\rangle= \langle N\rangle^2\langle p_i\rangle\langle p_j\rangle$ to obtain ${\cal C}_{ij}$. 
%
%Our physics arguments suggest that shear modes $r^{ij}_G$ in (\ref{eq:RGdef}) drive the correlations of ${\cal{M}}$; we prove this shortly. 
%For now, we identify  $\int \langle {\cal M}_i(\eta_1) {\cal M}_j(\eta_2)\rangle d\eta_r d\eta_a = \int \langle r^{ij}_G \rangle d\eta_r d\eta_a$, so that
%%
%\begin{eqnarray}\label{unres}
% \int \langle r^{ij}_G\rangle d\eta_r d\eta_a
% &=&\langle \sum_{{\rm all}\, a,b} p_{i,a}p_{j,b}\rangle - \langle N\rangle \langle p_i\rangle \langle  p_j\rangle
% \nonumber\\
%    &=&\langle N\rangle^2{\cal C}_{ij}+\langle\sum_{a} p_{i,a}p_{j,a}\rangle;
%\end{eqnarray}
%%
%the second equality follows from (\ref{Cdef}). 
%
Combining (\ref{unres0}) and (\ref{unres1}) then gives
\begin{eqnarray}\label{unres4}
\langle N\rangle^2{\cal C}_{ij}  
 &=& \int \langle r^{ij}_G\rangle d\eta_r d\eta_a \nonumber\\&&\;\;\;\; +\; {\rm cov}(P_i,P_j) - \langle \sum_{a} p_{i,a}p_{j,a}\rangle,
 %\langle N\rangle \langle p_i\rangle \langle  p_j\rangle
\end{eqnarray}
where ${\rm cov}(P_i, P_j) = \langle P_iP_j\rangle - \langle P_i\rangle\langle P_j\rangle$.
In local equilibrium
${\cal C}_{ij}\equiv 0$, so that 
\begin{equation}\label{unres2}
\int \langle r^{ij}_{G,\,\rm le}\rangle d\eta_r d\eta_a   =
\langle\sum_{a} p_{i,a}p_{j,a}\rangle-{\rm cov}(P_i,P_j).
%\langle P_iP_j\rangle + \langle P_i\rangle\langle P_j\rangle.
\end{equation}
Subtracting (\ref{unres2}) from (\ref{unres0}) and using (\ref{unres1}) gives (\ref{CcorrF}).
 
We comment that the ${\rm cov}(P_i,P_j)$ term on the second line of (\ref{unres4}) represents fluctuations of the total momentum in the rapidity interval from event to event. The second term includes additional fluctuations from the noise in each event.  %This term can be computed by integrating the $r^{ij}_{g,\,\rm le}$ obtained in deriving (\ref{eq:NoiseMom}). 

Generally, $\mathbf{M}$ combines shear flow $\mathbf{g}$ with a curl-free contribution, $\mathbf{g}_{l}$.  However, $\mathbf{g}_{l}$ does not contribute to the integral quantity ${\cal{M}}_i$, because we can write  $\mathbf{g}_{l} =\bm{\nabla} \varphi$. The rapidity density  ${\cal{M}}_i$ is then proportional to $\int  dx_i \partial \varphi/\partial x_i$, which depends only on the value of the potential $\varphi$ on the spatial part of the freeze out surface, where interactions effectively cease. There is no resorting force for ripples in this surface as there would be, e.g., for ocean waves. The curl-free contribution $\mathbf{g}_{l}$ to fluctuations at the freeze out surface must therefore be along the normal direction, so that the surface is an equipotential. The net contribution of $\mathbf{g}_{l}$ to ${\cal{M}}_i$ therefore vanishes. 

Observe that (\ref{Cfull}) implies a fixed value for the integral of $\langle\Delta r^{ij}_G\rangle$ over all rapidity when all particles are measured. A system completely constrained by momentum conservation can never reach the uncorrelated local equilibrium state.  Mathematically, this constraint constitutes a boundary condition for $\langle\Delta r^{ij}_G\rangle$ that amounts to a rapidity independent shift in magnitude. 

Experimental studies of momentum correlations have focused on $p_t$, rather than $p_x$ and $p_y$. In ref.\ \cite{Gavin:2006xd} we advocated studying such fluctuations using
\begin{eqnarray}\label{C0def}
   {\cal C} &=&  \langle N\rangle^{-2}\langle \sum_{a\neq b} p_{t,a}p_{t,b}\rangle 
   -\langle p_t\rangle^2 \nonumber\\
  &=& \langle N\rangle^{-2}\int  \langle \Delta r_G(\eta_r, \eta_a)\rangle
   d\eta_r d\eta_a,  
\end{eqnarray}
where $\langle \Delta r_G\rangle$ is the rapidity correlation function for the density $G = \tau\!\int g_r  rdrd\phi$, where $g_r$ is the radial component. Most of the basic arguments relating the rapidity dependence of ${\cal C}$ to the corresponding correlation function $\Delta r$ follow as above. The difference is that $p_t$ is not a conserved quantity. For all particles in the full rapidity range, (\ref{Cfull}) is replaced by
\begin{equation}\label{CTfull}
   {\cal C}\rightarrow  \frac{\langle (P_t - \langle P_t\rangle)^2\rangle}{\langle N\rangle^2}-\frac{\langle p_t^2\rangle}{\langle N\rangle}; %,\,\,\,\,\,\,\,\,\,\,\,\,\,\,{\rm all~particles}
\end{equation}
the fluctuations of total $P_t$ from event to event can be quite large and dependent on experimental details.

The STAR collaboration at RHIC reports a differential version of the quantity $\cal C$ as a function of relative pseudorapidity $\eta_r$ and azimuthal angle $\phi_r$ of pairs: 
\begin{equation}\label{exptC}
{\cal C}(\eta_r, \phi_r) =
\frac{\left\langle \sum\limits_{a\neq b} p_{_{t,a}}p_{_{t,b}}\right\rangle_{1,2}}
{\langle N \rangle_1 \langle N \rangle_2}
- \langle p_t\rangle_1 \langle p_t\rangle_2,
\end{equation}
where the numbers $\langle N \rangle_k$ and $\langle p_t \rangle_k$ refer to the particle number and transverse momentum in $(\eta_k, \phi_k)$ bins for particles $k= 1,2$ \cite{Agakishiev:2011fs}. The broad features of the two particle correlations displayed by (\ref{exptC}) as functions of $\eta_r$ and $\phi_r$ are quite familiar from measurements that omit the momentum weights.  The differential ${\cal C}(\eta_r, \phi_r)$ shows the usual ridge near $\phi_r = 0$ as a function of $\eta_r$. This near-side structure builds to a large symmetric peak at $\eta_r = 0, \phi_r = 0$. The away-side region also shows also a ridge centered about $\phi_r = \pi$ that is not as high and roughly independent of rapidity. 

The rapidity dependence of $\cal C$ is characterized by the width $\sigma$ of the near-side peak in $\eta_r$. In Au+Au collisions at the top RHIC energy, experimenters find that $\sigma$ increases from $0.54\pm 0.02 {\rm (statistical)} \pm 0.06 {\rm (systematic)}$ in the most peripheral collisions to $0.94\pm 0.06 {\rm (statistical)} \pm 0.17 {\rm (systematic)}$ in central collisions, consistent with predictions from ref.\ \cite{Gavin:2006xd} with a mean viscosity $\eta/s = 0.13\pm 0.03$. Significantly, STAR also presented the detailed rapidity distributions ${\cal C}(\eta_r)$ for a three centralities \cite{Agakishiev:2011fs} and for several other centralities \cite{PrivComm}.  We will study these measurements later.

 %***********************************************************************************************************************
%
%
\section{\label{sec:constNu}Diffusion vs. Experiment}
In this section we explore the behavior of $\Delta r_G$ and its influence on the qualitative features of $\cal{C}$. To keep our discussion here as simple as possible, we take $\tau_\pi^*$ and $\nu^*$ to be constant. Generally, to solve (\ref{eq:2VisDiff}) for $\Delta r_G$ we must first determine the behavior of the event-averaged temperature $T$ as a function of proper time using (\ref{eq:entdiff}), (\ref{eq:entheat}), and a realistic equation of state. The temperature then influences the evolution of fluctuations by changing the kinematic viscosity $\nu=\eta/Ts$, relaxation time $\tau_\pi = \beta \nu$, and the coefficient $\kappa$. This behavior is important for a quantitative analysis, but it makes systematic understanding of the equations very difficult. Taking constant $\tau_\pi^*$ and $\nu^*$ decouples (\ref{eq:2VisDiff}) from (\ref{eq:entdiff}) and (\ref{eq:entheat}). Furthermore, with this assumption we need not distinguish event and thermal averages. We therefore drop the brackets around $\Delta r_G$. We will study more realistic transport coefficients in future work. 

The most important feature of $\Delta r_G$ is its width in relative rapidity. Identified as an observable sensitive to viscosity in ref.\ \cite{Gavin:2006xd}, this width has since been measured \cite{Agakishiev:2011fs}. To compute the width, we follow ref.\ \cite{Aziz:2004qu} and multiply (\ref{eq:2VisDiff}) by $\eta_r^n$. Next, we integrate over $\eta_r$ and $\eta_a$ and use $\int\eta_r^n\partial^2\Delta r_G/\partial\eta_r^2 = n(n-1)\int\eta_r^{n-2}\Delta r_G$, which is nonzero only for $n \ge 2$.  We find 
\begin{equation}\label{eq:moments}
\left( \frac{\tau_\pi^*}{2}\frac{d^2}{d\tau^2} + \frac{d}{d\tau}\right)
A\langle\eta_r^n\rangle =
\frac{2\nu^*}{\tau^2}n(n-1)A\langle\eta_r^{n-2}\rangle, 
\end{equation}
where $\langle\eta_r^n\rangle = A^{-1}\int\eta_r^n\Delta r_G d\eta_r d\eta_a$ are the normalized moments of the rapidity correlation function. The amplitude $A$ and the mean $\langle \eta_r\rangle$ both satisfy (\ref{eq:moments}) with the  right side equal to zero. We take them to be constant and, moreover, take $\langle\eta_r\rangle=0$ assuming a symmetric system.  

The second moment gives the rapidity width $\sigma^2 = \langle\eta_r^2\rangle$, which satisfies
\begin{equation}\label{eq:moment2}
\left( \frac{\tau_\pi^*}{2}\frac{d^2}{d\tau^2} + \frac{d}{d\tau}\right)
\sigma^2 = \frac{4\nu^*}{\tau^2}.
\end{equation}
This equation holds generally for time and temperature dependent $\nu^*$ and $\tau_\pi^*$.  However, with constant values of these parameters, we see that the increase of the width is a function of the lifetime of the system alone.  

First order diffusion is described by (\ref{eq:moment2}) for $\tau_\pi^* = 0$ and $\nu^* = \nu$. We solve (\ref{eq:moment2}) for constant $\nu$ to find
\begin{equation}\label{eq:DeltaV}
\sigma^2 =\sigma_0^2+ \frac{4\nu}{\tau_0}\left(1-\frac{\tau_0}{\tau}\right),
\end{equation}
a result first obtained in ref.\ \cite{Gavin:2006xd}. 
Diffusion increases the width quickly and a-causally at early times, reaching the asymptotic value 
\begin{equation}\label{eq:DeltaVinf}
\sigma_\infty^2=\sigma_0^2 + 4\nu/\tau_0.
\end{equation}
%First order dynamics drives the width to this value very rapidly, with a time scale set by the formation time $\tau_0$. 
%
This saturation of the rapidity width to the value (\ref{eq:DeltaVinf}) is a straightforward consequence of Bjorken flow. In a stationary liquid, a spike in momentum diffuses over a range $\sim (2\nu t)^{1/2}$ that grows with time $t$.  Bjorken expansion of the underlying fluid stretches the longitudinal scale $\propto t$, rapidly overtaking diffusion and ``freezing in'' the initial inhomogeneity.      

%
%	FIGURE:		RAPIDITY WIDTH VS CENTRALITY
%
%
\begin{figure}%[hbt]% \begin{figure*} \end(figure*} * allows the figure to span the whole page
\includegraphics[width = \linewidth]{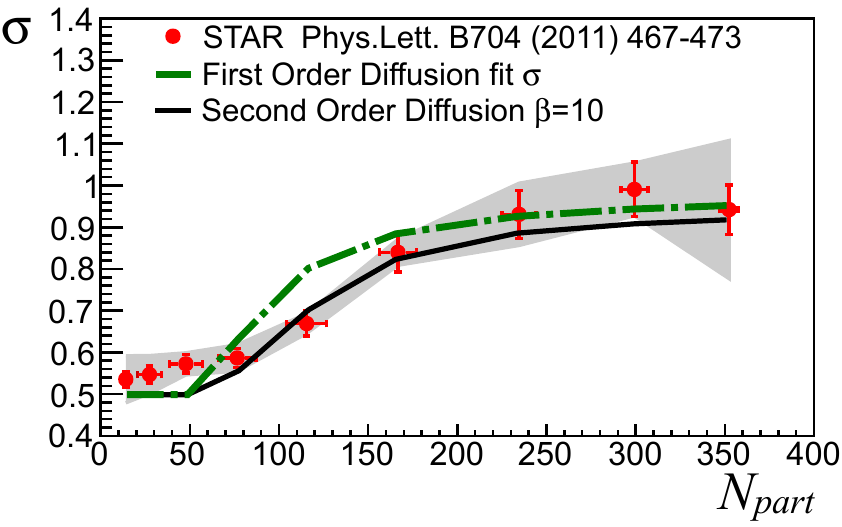}%
\caption{\label{fig:SigTau_constNu}(color online) 
Rapidity width as a function of the number of participants for second order momentum diffusion calculations (solid curve) compared to first order results. 
%Two schemes are used for the first order case: best fit to these data dash-dotted curves)
%and best fit to the distribution shape in fig.\ 2 (dashed).    
Data (solid circles) from STAR include shaded area to denote the systematic uncertainty in the fit procedure \cite{Agakishiev:2011fs}. 
}
\end{figure}
In fig.\ \ref{fig:SigTau_constNu} we show experimental measurements of the rapidity width of the near-side peak of the differential correlation function \cite{Agakishiev:2011fs}.  We present these results as a function of the number of participants $N_{part}$ to gauge the centrality. To compare first order diffusion to the measured widths (\ref{eq:DeltaVinf}), we must specify the freeze out time $\tau_F$ as a function of $N_{part}$. Hydrodynamic calculations with a hadronic afterburner are consistent with $\tau_F$ increasing roughly as a square of the root-mean-square radius of the participants $R$ \cite{Teaney:2001av}. We approximate that behavior as 
\begin{equation}\label{eq:tauF}
 \tau_F - \tau_0 = K(R(N_{part})-R_0)^2
\end{equation}
where $\tau_0$ is the formation time and $R_0$ is roughly the proton size. We compute $N_{part}$ and $R$ from a Glauber model and fix the constant $K$ so that the freeze out time in the most central collisions has a specified value $\tau_{Fc}$. 

%
%	FIGURE:		RAPIDITY CORRELATIONS 
%
%
\begin{figure*}[hbt]% \begin{figure*} \end(figure*} * allows the figure to span the whole page
\includegraphics[scale=1.0]{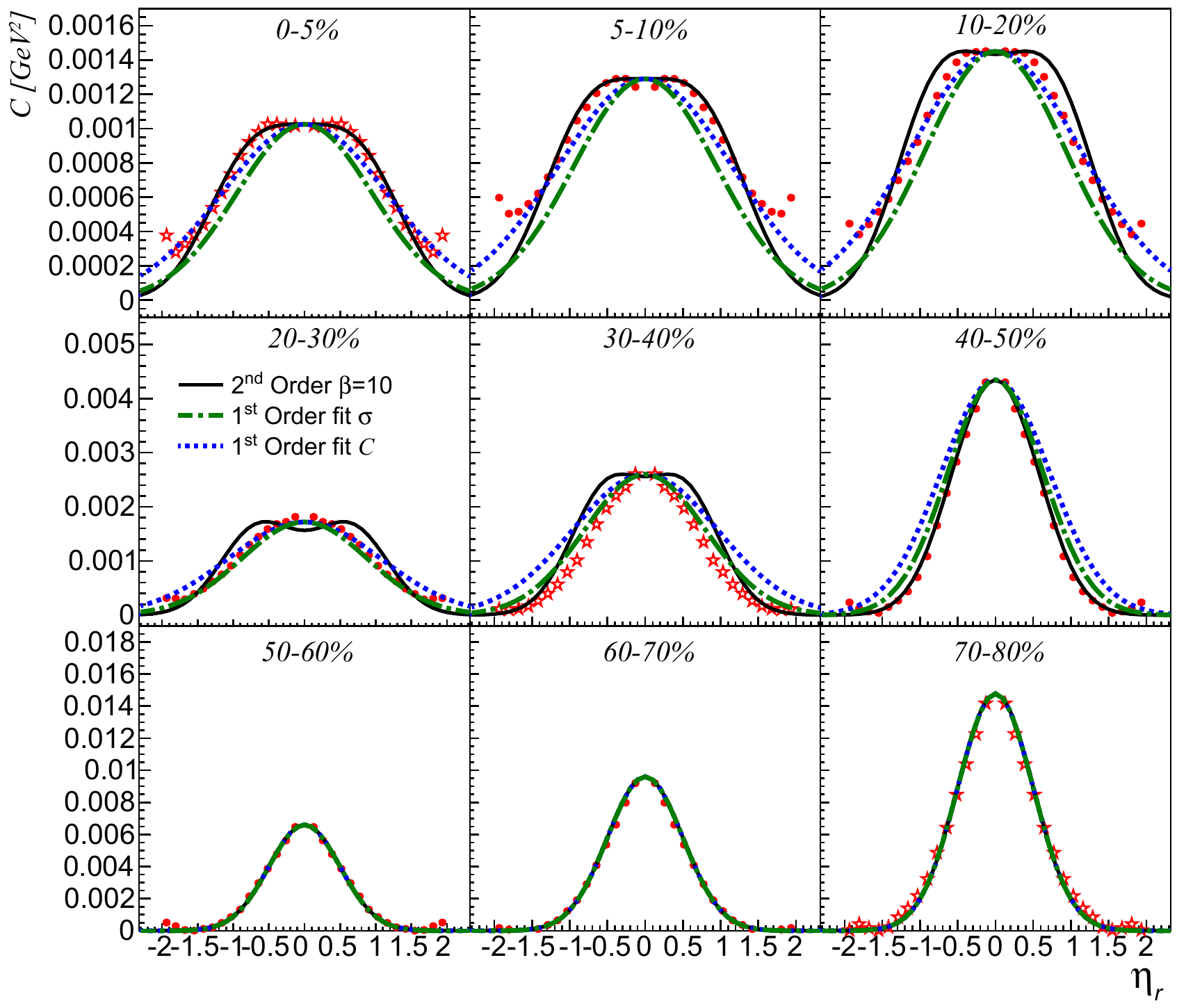}%
\caption{\label{fig:ProfileAll}(color online) 
Second order momentum diffusion calculations (solid curve)  compared to the rapidity dependence of the measured covariance (\ref{exptC}). First order calculations are also compared for best fit to these data (dashed)
and best fit to $\sigma$ in fig.\ \ref{fig:SigTau_constNu} (dash-dotted curves).    
Data (open stars) are from \cite{Agakishiev:2011fs} and (filled circles) from \cite{PrivComm}. Percentages of the cross section indicate centrality, with each panel corresponding to a width measurement in fig.\ \ref{fig:SigTau_constNu}. 
}
\end{figure*}
The rapidity width in first order diffusion rises with increasing centrality in rough accord with data, as shown in fig.\ \ref{fig:SigTau_constNu}. The dash-dot curve shows our best fit to this data using (\ref{eq:DeltaV}) evaluated at $\tau_F$, eq.\  (\ref{eq:tauF}).  Agreement depends mainly on the kinematic viscosity $\nu = \eta/Ts$, where $\eta/s = 1/4\pi$ and $T$ is the freeze out temperature.
Here we take $T = 140$~MeV to be the same for all centralities. Values of the space time parameters $\tau_0 = 0.65$~fm and $\tau_{Fc} = 12$~fm 
then specify (\ref{eq:DeltaVinf}) and the lifetime (\ref{eq:tauF}), respectively.   

Though overall agreement in fig.\ \ref{fig:SigTau_constNu} is adequate, our first order result is consistently above the data in the region where the data grows the most rapidly. This disagreement is due to the rapid rise of the width (\ref{eq:DeltaV}) with $\tau=\tau_F$ in first order diffusion. 

To find the rapidity width for second order diffusion, we solve (\ref{eq:moment2}) for constant $\tau_\pi^*=\tau_\pi$ and $\nu^* = \nu$. We must now specify an initial condition for $d\sigma^2/d\tau\equiv \theta_0^2$ at $\tau = \tau_0$, the value of which is unknown. An analogous situation arises when solving the one-body equations (\ref{eq:entdiff}) and (\ref{eq:entheat}), for which we must specify an initial value for $\Phi$. Some authors take $\Phi_0=4\eta/3\tau_0$, the Navier-Stokes value \cite{Dusling:2007gi,Song:2007ux}. This assumption aims to reduce the relative importance of second order corrections to Navier-Stokes behavior, as explained in ref.\ \cite{Dusling:2007gi}. In that spirit, we take the initial correlation function to satisfy 
\begin{equation}\label{eq:CausalIC}
\frac{\partial \Delta r_G}{\partial \tau}\Big|_{\tau=\tau_0}
= \frac{\nu_0}{\tau_0^2}\left( 2\frac{\partial^2}{\partial \eta_r^2} + \frac{1}{2}\frac{\partial^2}{\partial \eta_a^2} \right)
  \Delta r_G,
\end{equation}
corresponding to $\theta_0^2=4\nu/\tau_0^2$; see the discussion of (\ref{eq:2VisDiffRelax}) and (\ref{eq:2VisDiffRelax2}). In the absence of microscopic information on the initial conditions, this seems a reasonable choice. We consider an alternative ansatz  $\theta_0^2 = 0$ in the next section.  

Solving (\ref{eq:moment2}) we find
\begin{eqnarray}\label{eq:SOwidth}
\sigma^2 = \sigma_0^2 &+&
\frac{\theta_0^2 \tau_\pi}{2} \left(1 - e^{-2(\tau-\tau_0)/\tau_\pi}\right)\nonumber\\ 
&+& \frac{8\nu}{\tau_\pi}\int\limits_{\tau_0}^\tau \! du \!
\int\limits_{\tau_0}^u \! \frac{ds}{s^2} e^{2(s-u)/\tau_\pi}.
\end{eqnarray}
%
%\begin{equation}\label{eq:SOwidth}
%\sigma^2 = \sigma_0^2 +
%\frac{\theta_0^2 \tau_\pi}{2} \left(1 - e^{-2(\tau-\tau_0)/\tau_\pi}\right)+ \frac{4\nu}{\tau_0}F\left(\frac{\tau}{\tau_0}\right), 
%\end{equation}
%for  $\alpha = 2\tau_0/\tau_\pi$ and 
%\begin{equation}\label{eq:SOwidth2}
%F(s)= \alpha\int\limits_{1}^{s} \! du \!
%\int\limits_{1}^u \! dv e^{\alpha(v-u)}/v^2
%\end{equation}
%%
The solid black curve in fig.\ \ref{fig:SigTau_constNu} shows the value of (\ref{eq:SOwidth}) at the freeze out time (\ref{eq:tauF}) in comparison to the data. Again we take $\nu = \eta/Ts$ for $\eta/s = 1/4\pi$, but now with $T=150$~MeV for all centralities. We must now specify the second-order relaxation time $\tau_\pi = \beta \nu$, for which we take $\beta = 10$.  The values $\tau_0 = 1.0$~fm and $\tau_{Fc} = 10$~fm then give superb agreement with data. %We remark that calculations for a kinetic theory value $\beta = 5$ fall between the first and second order results.      

Observe that any solution of (\ref{eq:moment2}) reaches a ``terminal velocity''  $d\sigma^2/d\tau = 4\nu/\tau^2$ for $\tau \gg \tau_\pi$, so that $\sigma^2$ approaches the first order result (\ref{eq:DeltaV}) plus a constant.  For $\theta_0^2=4\nu/\tau_0^2$, the width approaches the asymptotic value 
\begin{equation}\label{eq:2DeltaVinf}
\sigma_\infty^2=\sigma_0^2 + \frac{4\nu}{\tau_0}\left(1+\frac{1}{2}\frac{\tau_\pi}{\tau_0}\right),
\end{equation}
which is larger than the first order limit (\ref{eq:DeltaV}). Consequently, different parameter values are needed for the first and second order fits in fig.\ \ref{fig:SigTau_constNu}. For $\theta_0^2 =0$ the solution approaches the first order value  (\ref{eq:DeltaV}) from below for $\tau\gg \tau_\pi$. 
We will come back to this point in the next section. 
 
To lay bare the difference between first and second order evolution, we turn to the shape of the differential correlation function ${\cal{C}}$ as a function of $\eta_r$; see eq.\ (\ref{exptC}). STAR reported ${\cal{C}}(\eta_r)$ for three centralities represented  as open stars in fig.\  \ref{fig:ProfileAll} \cite{Agakishiev:2011fs}. Additional centralities are shown as solid circles \cite{PrivComm}. Percentages labeling each panel indicate the centrality defined by the fraction of total cross section. Every panel in fig.\ \ref{fig:ProfileAll} corresponds to a width in fig.\ \ref{fig:SigTau_constNu}. Experimenters fit the near-side peak of the measured distributions with a double-Gaussian function plus a constant offset. They then subtracted the offset from the measured values to calculate the rapidity width in fig.\ \ref{fig:SigTau_constNu}. The error band here represents the uncertainty in this fit procedure. The measured ${\cal C}(\eta_r)$ are shown here with the offsets from ref.\ \cite{Agakishiev:2011fs,PrivComm} subtracted.

We now solve (\ref{eq:2VisDiff}) to compute the correlation function $\Delta r_{G}$ and its integral ${\cal{C}}(\eta_r)$, assuming the initial transverse momentum correlation function to be
\begin{equation}\label{eq:Initial_rg}
\Delta r_G(\eta_r,\eta_a,\tau_0) = A e^{-\eta_r^2/2\sigma_0^2}e^{-\eta_a^2/2\Sigma_0^2}.
%\frac{e^{-\eta_r^2/2\sigma_0^2}e^{-\eta_a^2/2\Sigma_0^2}}{2\pi\sigma_0\Sigma_0}.
\end{equation}
This distribution is motivated by the rapidity dependence of measured correlation functions for multiplicity and net charge in pp collisions. We set the initial width in relative rapidity, $\sigma_0$ to fit the most peripheral distribution in fig.\ \ref{fig:ProfileAll}. Furthermore, we assume there is insufficient time for significant evolution in the three most peripheral cases. The data supports this claim and give a consistent value of $\sigma_0 = 0.50$. The average pseudo-rapidity width $\Sigma_0\approx 5-6$ units is assumed to be a ``large'' value relative to the size of experimental acceptance. 
We will take $A$ to fit the peak value of the measured $\cal{C}$. This parameter has little impact on our current study, since we are only concerned with the shape of the function.  We use (\ref{eq:CausalIC}) for the initial value of the first derivative.

%We comment that it is not straightforward to compare (\ref{C0def}) to the amplitude of the measured differential distribution (\ref{exptC}) because the 
%latter $\phi$-dependent quantity is affected by azimuthally anisotropic flow.  Arguments in ref.\ \cite{Gavin:2011gr} show that the contribution of anisotropy to the $\phi$ integrated quantity (\ref{CTfull}) is suppressed. In essence, $\cal{C}$ adds the scalar $p_t$ of particles irrespective of their $\phi$ direction. % From our point of view, this is an important motivation for measuring these and similar integrated quantities.  
%On the other hand, it is evident from data in \cite{Agakishiev:2011fs} and simulations in \cite{Sharma:2011nj} that anisotropic flow influences the away-side behavior of the differential distribution (\ref{exptC}) and, by inference, the near-side. Anisotropic flow is largely a long range correlation that varies slowly with rapidity. Flow effects are likely removed when experimenters subtract their rapidity independent offset.  

First order momentum diffusion yields a Gaussian rapidity profile. For $\tau_\pi^* = 0$, (\ref{eq:2VisDiff}) reduces to (\ref{eq:2VisDiffRelax2}).  Evolution preserves the Gaussian initial shape (\ref{eq:Initial_rg}), so that integration over $\eta_a$ yields a Gaussian in $\eta_r$ of width (\ref{eq:2DeltaVinf}).  

We find first order momentum diffusion to be inconsistent with the measurements in fig.\  \ref{fig:ProfileAll}, despite overall agreement with the width in fig.\ \ref{fig:SigTau_constNu}. Our most reliable first order calculations give the dash-dot curves in fig.\ \ref{fig:SigTau_constNu} and fig.\  \ref{fig:ProfileAll}.  We adjust the parameters to obtain best agreement with the rapidity width data in fig.\ \ref{fig:SigTau_constNu}, and predict the rapidity shape in fig.\ \ref{fig:ProfileAll}. These calculations fail miserably to describe the measured rapidity profiles.  We are confident in this fit because these width measurements were the focus of the experimental study, so that systematic errors were provided. We next ask whether first order diffusion can be brought closer to agreement with the rapidity shape by fitting the data in fig.\ \ref{fig:ProfileAll} alone. The dashed curves in fig.\ \ref{fig:ProfileAll} are computed for parameter values $\eta/s = 1/4\pi$, $T = 110$~MeV, $\tau_0 = 0.50$~fm, and $\tau_{Fc} = 10$~fm. Agreement with the measured shape is still quite poor.      

The measured distributions in the top three panels of  fig.\ \ref{fig:ProfileAll} differ from the Gaussian profile of first order diffusion in two telling ways. First, they are systematically broader, with a flatter peak.  Second, they show a small dip near $\eta_r = 0$, suggesting a bimodal nature.  The flattening feature is the most compelling -- this is why first order diffusion fails. Furthermore, we consider the bimodal feature an intriguing possibility. Several points in the 0-5\% and 5-10\% panels indicate double peaks. Note that the experimenters omit $\eta_r = 0$ bins appearing in \cite{Agakishiev:2011fs}, as they are fraught with track-merging and other experimental challenges \cite{PrivComm}. The experimenters also took this bimodal structure seriously, fitting their data as a double-Gaussian function plus a constant offset \cite{Agakishiev:2011fs}, a result that first order diffusion can never generate. 

Is the bimodal nature of the data a consequence of second order evolution? Causal diffusion broadens the rapidity distribution by wave-like propagation of the initial signal in addition to the usual diffusion. Mathematically, the $\tau_\pi^*$ term in (\ref{eq:2VisDiff}) changes the equation from parabolic to hyperbolic, like a wave equation. In wave motion, a Gaussian initial pulse divides into half-amplitude pulses propagating to the right and left in the $z$ coordinate at wave speed $v$. In (\ref{eq:bj4}) the speed is $v=\sqrt{\nu/\tau_\pi}$. Observe that the wave speed diverges as $\tau_\pi \rightarrow 0$ and we approach the first order diffusion regime, thus violating causality.  In rapidity coordinates, this separation is less pronounced because rapidity measures speed $z/t$, not position. 

%
%	FIGURE:		RAPIDITY PROFILE VS TAU
%
%
\begin{figure}%[hbt]% \begin{figure*} \end(figure*} * allows the figure to span the whole page
\includegraphics[scale=1.0]{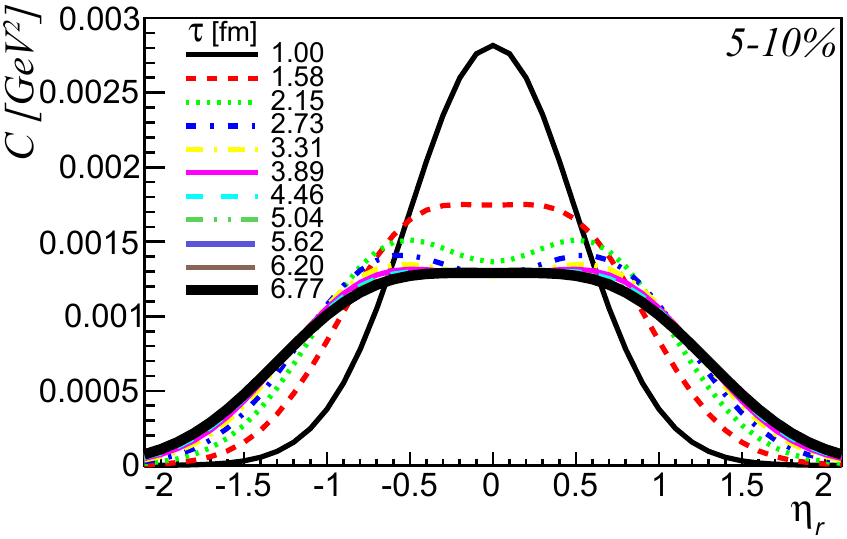}%
\caption{\label{fig:fig3}(color online) 
Time dependence of the rapidity covariance in second order diffusion.  
}
\end{figure}
The time evolution of the rapidity profile is shown in fig.\ \ref{fig:fig3} for parameter values used in fig.\  \ref{fig:ProfileAll}. In the 5-10\% centrality range shown, evolution starts at $\tau_0 = 1.0$~fm and ends at 6.8~fm. The evolution is initially wavelike, giving rise to left and right moving pulses. After a time $\sim \tau_\pi$ has elapsed, the first derivative in the left side of (\ref{eq:2VisDiff}) becomes important and diffusion begins. This diffusion works to fill in the gap between the pulses and create a single broad plateau over a time $\sim \sigma^2\tau^2/\nu$. The $\tau^{2}$ factor, which comes from the right side of (\ref{eq:2VisDiff}), eventually slows diffusion to an extent the rapidity profile becomes ``frozen.''  How far this evolution can progress for collisions in a given centrality class depends on the freeze out time (\ref{eq:tauF}) compared to these other time scales.  Whether distinct peaks can be resolved further depends on the pulse width $\sim \sigma_0$ compared to the asymptotic increase $\sigma_\infty - \sigma_0$, given by (\ref{eq:2DeltaVinf}).  

Our solution of  (\ref{eq:2VisDiff}) gives the solid curves in fig.\ \ref{fig:ProfileAll}. The evolution from peripheral to central reflects the time evolution in fig.\ \ref{fig:fig3} due to the increase of $\tau_F$ described by (\ref{eq:tauF}).  Our calculations agree very well with the measured shape rapidity profiles for the three most central distributions. They also agree with the widths in fig.\ \ref{fig:SigTau_constNu} for all centralities.

We emphasize that the evolution of the rapidity landscape from a single peak in peripheral collisions to a broader plateau for more central collisions is characteristic of second order diffusion. Second order calculations with the initial condition (\ref{eq:CausalIC}) show this behavior very strongly. For constant $\tau_\pi^*$ and $\nu^*$ centrality dependence is solely determined by $\tau_F$, (\ref{eq:tauF}). With temperature and time dependent parameters, further complexity follows from the dependence on the initial temperature and features of the equation of state.   

%Our neglect of average transverse flow is inspired by NeXSPheRIO simulations in ref.\  \cite{Sharma:2011nj}, which demonstrate that the average flow does not appreciably alter the distribution of fluctuations ${\cal C}(\eta_r)$.  

We now comment on the effect of transverse flow on these phenomena. NeXSPheRIO simulations in ref.\  \cite{Sharma:2011nj} demonstrate that the average transverse flow does not appreciably alter the rapidity distribution of its fluctuations, ${\cal C}(\eta_r)$.  These event-by-event  hydrodynamic simulations are broadly consistent with the azimuthal-angular dependence of two particle correlations. Nevertheless, this code omits viscosity and thermal fluctuations, so we would not expect it to describe the changes in ${\cal C}(\eta_r)$ that we discuss in this section. Indeed, NeXSPheRIO simulations are essentially Gaussian for all centralities \cite{Sharma:2011nj}. The rapidity width does not  increase with centrality, nor does the shape of ${\cal C}(\eta_r)$ change. This result supports our neglect of mean transverse flow in this paper.  Furthermore, it  fortifies our interpretation of the data as consequences of second order viscous dissipation.  %The simulated width in $\eta_r$ may possibly decrease by $\sim 10\%$ -- albeit with large uncertainties -- in agreement with our expectation that transverse flow can only focus particles into a narrower angular region. We can be confident in our computation of the width in fig.\ \ref{fig:SigTau_constNu} at this $10\%$ level.  

A further consequence of transverse flow is that it generates the azimuthal anisotropy of flow. This anisotropy causes the difference 
$\gamma^\prime \propto ({\cal C}_{yy} - {\cal C}_{xx})$ \cite{Pratt:2010zn}.  It is reasonable to ask what effect this anisotropy might have on the $p_t$ covariance (\ref{C0def}) and the near-side amplitude of the differential distribution (\ref{exptC}) as measured. A key motivation in ref.\ \cite{Gavin:2006xd} was to find a measure of viscosity that is independent of this anisotropy.  Arguments in ref.\ \cite{Gavin:2011gr} show that the contribution of anisotropy to the $\phi$ integrated quantity (\ref{CTfull}) is suppressed. In essence, $\cal{C}$ adds the scalar $p_t$ of particles irrespective of their $\phi$ direction. On the other hand, it is evident from data in \cite{Agakishiev:2011fs} and simulations in \cite{Sharma:2011nj} that anisotropic flow influences the away-side behavior of the differential distribution (\ref{exptC}) and, by inference, the near-side. Nevertheless, anisotropic flow is largely a long range correlation that varies slowly with rapidity. Flow effects are likely removed when experimenters subtract their rapidity independent offset.

%***********************************************************************************************************************
%
%
\section{\label{sec:betaPlots} How to Measure $\tau_\pi$}
In the spirit of ref.\ \cite{Gavin:2006xd} we now ask how one can measure the second-order transport coefficient $\tau_\pi$. Most work on measuring transport coefficients in nuclear collisions has focused on extracting $\eta/s$ from azimuthal anisotropy measurements.  Niemi et al.\  found that changes in $\eta/s$ could be compensated by changing $\tau_\pi$ to yield the same anisotropic flow \cite{Niemi:2012ry}. To vary $\tau_\pi$ with everything else fixed, one writes 
\begin{equation}\label{eq:betaDef}
\tau_\pi = \beta\nu 
\end{equation}
%
%$\tau_\pi = \beta\nu$ 
and varies $\beta$. This form is inspired by kinetic theory, which gives $\beta = 5$ for massless particles obeying Boltzmann statistics. While causality requires $\beta \geq 2$, little else is known about its value \cite{Denicol:2011fa,Pu:2009fj}. Reference \cite{Niemi:2012ry} showed that the values  $\eta/s = 0.16$ and $\beta = 10$ yield practically the same $v_2$ coefficient as $\eta/s = 0.08$ and $\beta = 5$.  How then can we disentangle these contributions? 
%
%	FIGURE:		RAPIDITY WIDTH VS CENTRALITY
%
%
\begin{figure}% \begin{figure*} \end(figure*} * allows the figure to span the whole page
\includegraphics[width = \linewidth]{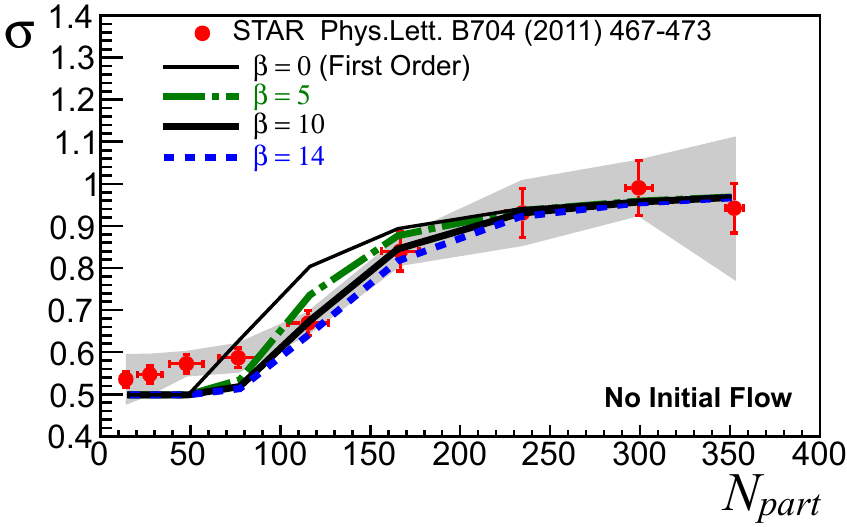}%
\caption{\label{fig:SigTauZero}(color online) 
The sensitivity of the rapidity width to the second order relation time $\tau_\pi$ illustrated using initial conditions with no initial flow. The different values of $\beta$ change $\tau_\pi = \beta\nu$ for fixed kinematic viscosity $\nu$ relative to first order $\beta =0$. Data is the same as in fig.\ \ref{fig:ProfileAll}. 
}
\end{figure}
%
%
%
%	FIGURE:		PROFILE for VARIOUS BETA VALUES
%
%
\begin{figure}% \begin{figure*} \end(figure*} * allows the figure to span the whole page
\includegraphics[scale=1.0]{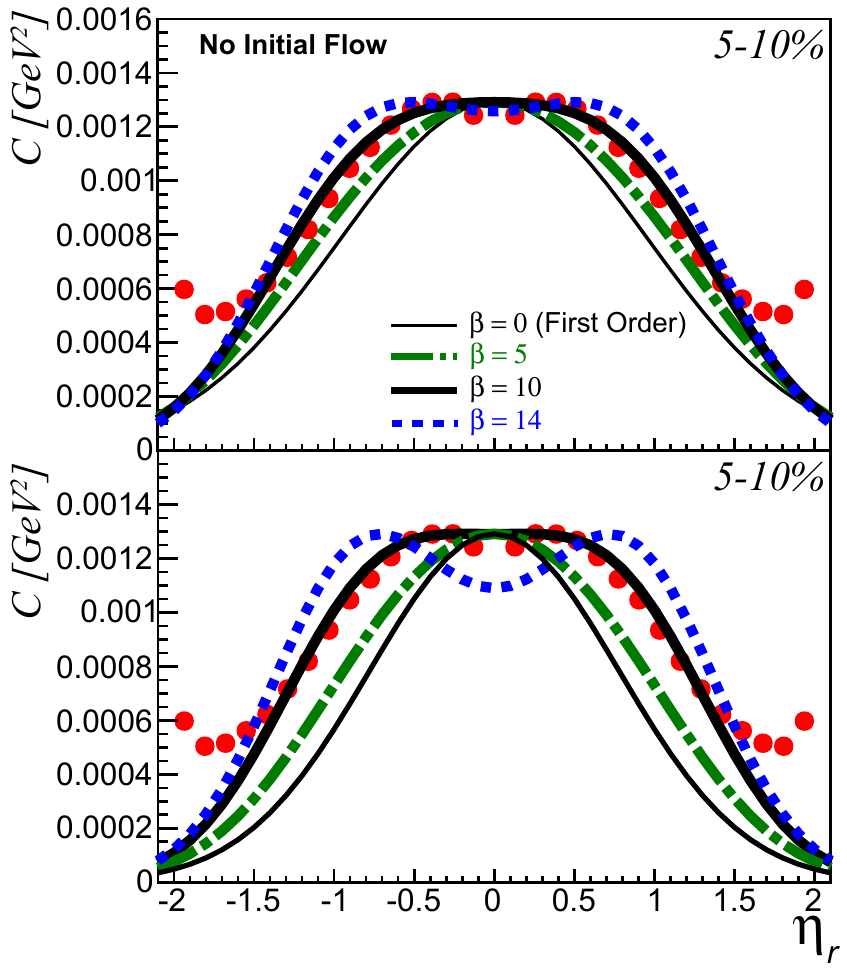}%
\caption{\label{fig:ChangeBeta}(color online) 
The sensitivity of the rapidity profile of  correlations to the second order relation time $\tau_\pi$. The different values of $\beta$ change $\tau_\pi = \beta\nu$ for fixed kinematic viscosity $\nu$ relative to first order $\beta =0$. The top panel uses the no-flow initial condition (\ref{eq:ZeroIC}), so that each curve has the same integrated width $\sigma$. The bottom panel uses near-equilibrium initial conditions (\ref{eq:CausalIC}), with $\sigma$ that follows (\ref{eq:2DeltaVinf}).
}
\end{figure}
%
%
%
%	FIGURE:		RAPIDITY CORRELATIONS ZERO IC
%
%
\begin{figure*}% \begin{figure*} \end(figure*} * allows the figure to span the whole page
\includegraphics[scale=1.0]{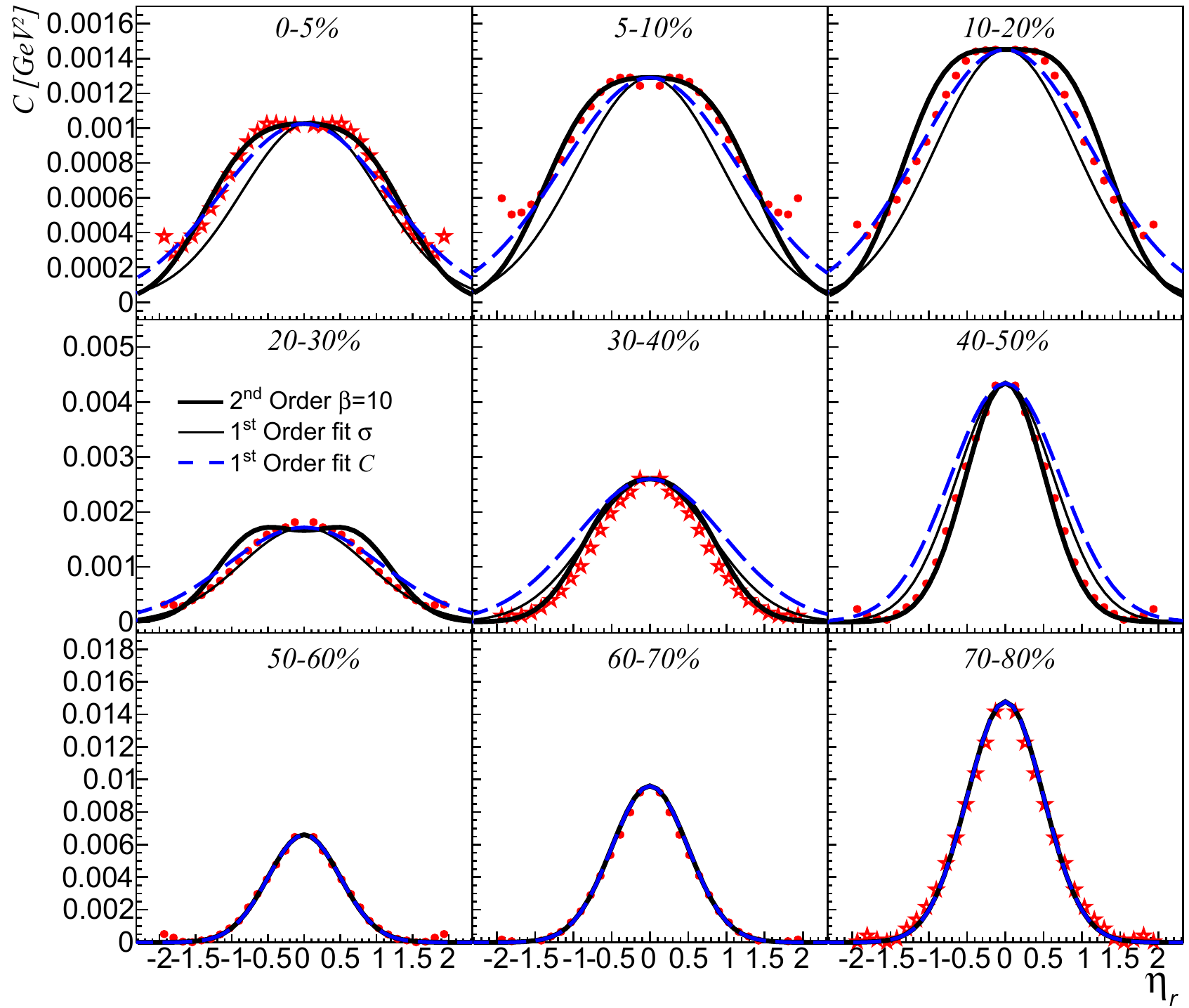}%
\caption{\label{fig:ProfileAllZero}(color online) 
Measured rapidity profile compared to the characteristic evolution of second order diffusion from single peak to plateau.   
Same as fig.\ \ref{fig:ProfileAll} but with no initial flow. 
}
\end{figure*}

The signature role of $\tau_\pi$ is in determining the rate at which the system relaxes to Navier-Stokes hydrodynamics. 
%To see how the rapidity behavior of the momentum covariance $\cal C$ can be used to measure $\tau_\pi$, observe that its signature effect is that it governs the rate at which the system relaxes to an asymptotic state described by Navier-Stokes hydrodynamics. 
We have seen two consequences of a finite $\tau_\pi$ in the previous section: First, the evolution changes from wavelike to diffusion-dominated, as illustrated in fig.\ \ref{fig:fig3}. Second, it modifies the flow that drives the increase of $\sigma$ toward the asymptotic value (\ref{eq:2DeltaVinf}). The first effect is uniquely a second order transport phenomenon governed by $\tau_\pi$. In contrast, viscous diffusion is at the heart of the second effect \cite{Gavin:2006xd}. The growth of $\sigma$ only acquires a $\tau_\pi$ dependence due to initial flow $\theta_0^2\equiv d(\sigma^2)/d\tau|_{\tau_0} \propto \nu/\tau_0^2$. %Details here are bound to change as our model becomes more sophisticated.  

To isolate the second order relaxation effect of $\tau_\pi$ and identify its consequences, we replace (\ref{eq:CausalIC}) with
\begin{equation}\label{eq:ZeroIC}
(\partial \Delta r_G/\partial \tau)|_{\tau=\tau_0} =  0,% \,\,\,\,\,\,\,\,\,\,\,\,\,\,{\rm no~initial~flow}
\end{equation}
corresponding to no initial flow, $\theta_0^2=0$.  While used here for illustrative purposes, such a non-equilibrium initial condition might be physically relevant if, e.g., the values of $\mathbf{g}$ and $\partial \mathbf{g}/\partial t$ are uncorrelated everywhere in each event in the initial state; see eq.\ (\ref{eq:2DiffusionGG}).  
This is analogous to taking the initial $\Phi=0$ when solving one-body equations (\ref{eq:entdiff}) and (\ref{eq:entheat}), another common choice among practitioners \cite{Niemi:2012ry}. 

We first compute the centrality dependence of the rapidity width using (\ref{eq:SOwidth}) and (\ref{eq:ZeroIC}). 
%To vary $\tau_\pi$ with everything else fixed, we write $\tau_\pi = \beta\nu$ and vary $\beta$. 
The results in fig.\ \ref{fig:SigTauZero} are then computed with $\eta/s = 1/4\pi$, $T = 143$~MeV, $\tau_0 = 0.6$~fm and $\tau_{Fc} = 10$~fm. 
The difference between figs.\ 1 and \ref{fig:SigTauZero} is striking.  The width calculated using  (\ref{eq:CausalIC})  asymptotically approaches $\sigma_\infty^2 = \sigma_0^2 + 4\nu/\tau_0$, the first order value (\ref{eq:DeltaVinf}). In contrast, $\sigma$ in fig.\ 1 includes initial flow that leads to the $\tau_\pi$-dependent asymptotic value (\ref{eq:2DeltaVinf}). This is a large effect in practice:  in fig.\ \ref{fig:SigTauZero} we compare first and second order calculations for the same parameter values, while this is impossible in fig.\ \ref{fig:SigTau_constNu}.  

Relaxation is the only effect of $\tau_\pi$ in fig.\ \ref{fig:SigTauZero}. Its impact is most evident where $\sigma$ increases most rapidly. The same effect is also evident in fig.\ \ref{fig:SigTau_constNu}, albeit convoluted with the increase in the asymptotic width.  Sufficiently precise measurements of the centrality dependence can yield information on $\tau_\pi$. While we find best agreement in both figs.\ \ref{fig:SigTau_constNu} and \ref{fig:SigTauZero} for $\beta = 10$, we hesitate to draw such quantitative conclusions from the present schematic calculation.   

%The broad shoulders and slight bimodal shape of the rapidity profile signifies the wavelike nature of second order diffusion. 
We exhibit the sensitivity of the rapidity profile to $\tau_\pi=\beta \nu$ in fig.\ \ref{fig:ChangeBeta}. The broadness of the shoulders compared to a smoothly sloping Gaussian is fully evident in the data, but the most interesting feature is the valley near $\eta_r = 0$, as it may indicate wavelike structure.  The top panel shows computations for no initial flow. Two things happen as we increase $\beta$. First, more time is available for wavelike structure to develop. Second, the system reaches the first order regime more slowly, reducing the time during which diffusion can fill the valley between the bumps. The width in this figure is constant, compensated by changes in the tails outside the plotting range.  

The profile computed with initial longitudinal flow (\ref{eq:CausalIC}) is shown in the bottom panel in  fig.\ \ref{fig:ChangeBeta}. In addition to the changes described earlier, the overall width of the curve grows as $\beta$ increases due to the increase of the asymptotic width $\sigma_\infty$, described by (\ref{eq:2DeltaVinf}).  As with the widths, the data favor a value $\beta\approx 10$ for both initial conditions. 
It is interesting that we are better able to resolve two peaks for the initial flow ansatz (\ref{eq:CausalIC}) than for (\ref{eq:ZeroIC}). It is precisely the larger difference between $\sigma_\infty$ and the initial width $\sigma_0$ with (\ref{eq:CausalIC}) that allows us to better resolve these peaks.

%Nevertheless, this might seem counterintuitive because starting the system initially with Navier-Stokes diffusion was intended to minimize the effect of second order corrections.

In fig.\ \ref{fig:ProfileAllZero} we show the rapidity profiles for the entire experimental centrality range obtained for initial conditions with no initial flow, (\ref{eq:ZeroIC}). 
Figures \ref{fig:ProfileAll} and \ref{fig:ProfileAllZero} taken together support our contention that second order evolution can explain these profiles better than first order diffusion. Furthermore, the results are well described by $\beta = 10$ regardless of initial conditions. %The agreement with $\beta = 10$ is slightly better in the mid-peripheral range in fig.\  \ref{fig:ProfileAllZero}. %
 %For the current constant-coefficient model, the resolution of distinct modes depends only on the size of the increase in $\sigma$ relative to its initial value. 
%The detailed behavior seen here can be much more complex with temperature and time dependent coefficients.   

%The rapidity profiles for the entire centrality range for the initial conditions that without initial flow (\ref{eq:ZeroIC}) are very similar to those in  fig.\  \ref{fig:ProfileAll}. If anything, the agreement with $\beta = 10$ is slightly better in the mid-peripheral range. We stress that any such details will likely change when a more realistic equation of state and temperature dependent viscosity are included. 

We again point out that the data follow the characteristic pattern of second order evolution from a single peak to a broader plateau for increasingly central collisions. The bimodal nature at intermediate centralities is less evident for the no-initial-flow calculations in fig.\ \ref{fig:ProfileAllZero}, compared to the initial condition (\ref{eq:CausalIC}) in fig.\ \ref{fig:ProfileAll}. The initial conditions without this flow seem more in accord with data.  However, systematic uncertainties for the distributions are not available, so we cannot say anything precise about the bimodal character of the distributions \cite{Agakishiev:2011fs,PrivComm}. Moreover, details of calculations at this level may change when temperature and time dependent coefficients are included.   

It is important to emphasize that we expect the rapid changes in $\sigma$ to {\em coincide} with the most dramatic shape changes as centrality is varied, provided that $\tau_\pi$ is the driving factor. Theoretically,  the relaxation of $\sigma$ and the wave-to-diffusion transition are both due to a competition between the first and second time derivatives in (\ref{eq:2VisDiff}), since (\ref{eq:moment2}) is derived from (\ref{eq:2VisDiff}). How far this competition progresses for collisions in a given centrality range depends on $\tau_\pi$ compared to the freeze out time. Experimentally, the three data points exhibiting the most rapidly increasing $\sigma$ in figs.\ \ref{fig:SigTau_constNu} and \ref{fig:SigTauZero} are derived from the distributions in the middle panels in figs.\  \ref{fig:ProfileAll} and \ref{fig:ProfileAllZero}. One can see that the shape of the distributions -- and not just their widths -- changes most rapidly in these panels.

%
%Some things to notice about Fig.\ref{fig:SigNp_Omit3}:
%\begin{itemize}
%\item To cut out the ``wings'' we omit three data points on each tail of each graph in Fig.\ref{fig:C2ndOrd}. We re-calculate the RMS width from the data using this omission to obtain the open red circles.
%\item if we calculate the RMS width of our calculation of ${\cal C}(\eta_r)$ in Fig.\ref{fig:C2ndOrd} for the same range of the data we get nicer and more appropriate comparison to the data.
%\end{itemize}

%***********************************************************************************************************************
%
%
\section{\label{sec:conclusion}Conclusion}
Our goal in this paper is to identify the key physics issues probed by momentum correlations. In earlier work we suggested how such correlations could be used to study viscosity \cite{Gavin:2006xd,Pratt:2010zn}.  Here we propose a way to measure $\tau_\pi$. This measurement relies on the hyperbolic nature of the second order transport. Our analysis is constructed from several different pieces, each with its own challenges. 

Computing the correlations in secs.\ \ref{sec:Roadmap} and \ref{sec:ionCollisions} required the use of hydrodynamics with noise and dissipation. We derived (\ref{eq:2VisDiff}) and used it to compute the observable correlation functions.  
The derivations in sec.\ \ref{sec:Roadmap} were lengthy, but essential. In addition to momentum diffusion we discussed Brownian motion and particle number diffusion. Equation (\ref{eq:AbdelAzizEquation}) that leads to (\ref{eq:2VisDiff}) improved on our early exploratory work in ref.\ \cite{Pokharel:2013rka}, which in turn relied on the heuristic formulation of ref.\ \cite{Aziz:2004qu}. The factor of $1/2$ in the second order term in (\ref{eq:2VisDiff}) that is new to this work can be understood by comparison to Brownian motion. In both eqs.\ (\ref{eq:Brownian6}) and (\ref{eq:2VisDiffRelax}), fluctuations equilibrate on a time scale $\sim \tau_\pi/2$ --- half the time needed for the mean to relax. Particle diffusion is easier to understand and better known in the literature \cite{van2011stochastic,gardiner2004handbook}. All three problems have applications in nuclear collisions.  

In sec.\ \ref{sec:Modes} we discussed the hydrodynamic shear modes in first and second order theory. We worked to establish the connection between the observables ${\cal{C}}_{ij}$, ${\cal{C}}$, and the shear momentum current in secs. \ref{sec:ionCollisions} and \ref{sec:Observables}.  This connection is important because shear modes do not couple at linear order to the other modes. Consequently,  (\ref{eq:2VisDiff}) only depends on $\tau_\pi$, $\nu$ and $\kappa$, allowing us to use the systematic behavior of data to extract these parameters.  Other modes important for other observables exhibit more complex behavior \cite{Floerchinger:2015efa}. 

We are working to extend our methods to other modes in order to address a wider range of observables.  In particular, similar hyperbolic behavior can appear in net charge and net baryon diffusion. Diffusion of net charge and baryon number including hydrodynamic fluctuations has been studied by a number of authors \cite{Aziz:2004qu,Kapusta:2014aza,Kapusta:2014dja,Ling:2013ksb,Floerchinger:2015efa,Asakawa:2015ybt}.  To apply (\ref{eq:2VisDiff}) to these systems, we can replace $\eta/w$ by $D$ and $\tau_\pi$ with the relaxation time for particle diffusion $\tau_d$. Whether one sees the tell-tale features of relativistic diffusion amid the effects of hadronization  discussed in ref.\ \cite{Pratt:2015jsa} is an interesting question for future study. 

In sec.\ \ref{sec:Observables} we discussed the observables. The analysis in secs.\ \ref{sec:constNu} and \ref{sec:betaPlots} are based on measurements of $p_t$ correlations using the observable ${\cal{C}}$ recommended in \cite{Gavin:2006xd}.  The data exclude the Gaussian shape of first order calculations \cite{Agakishiev:2011fs,PrivComm}. The better agreement of the broader, flatter second-order results is encouraging, but we are aware that both measurements and computations can be improved with the goal of measuring the shape -- not just the width -- in mind.  Moreover, the observable most closely connected to the transverse momentum fluctuations are ${\cal{C}}_{ij}$, the covariance of Cartesian components of the transverse momenta \cite{Pratt:2010zn}. We hope that RHIC beam energy scan and LHC can measure both quantities. ATLAS and CMS offer broader rapidity coverage at LHC, but they also have a higher minimum $p_t$, which may affect the analysis. It would be especially interesting to see if thermalization effects appear in $pA$ as well as $AA$ measurements.  We are also working to understand the relationship between ${\cal{C}}$ and other observables of longitudinal correlations, e.g., \cite{Bzdak:2012tp,SoorajRadhakrishnanfortheATLAS:2015eqq}.

Data from refs.\ \cite{Agakishiev:2011fs,PrivComm} are in good accord with calculations assuming $\tau_\pi/\nu = \beta = 10$ for both sets of initial conditions we tried. This value is large but not inconsistent with azimuthal flow calculations \cite{Niemi:2012ry}. While it differs appreciably from the estimate $\beta = 5$ from kinetic theory of massless Boltzmann particles, so famously does the viscosity. For our values of $\nu$ we estimate $\tau_\pi$ in the range from 1.0 to 1.1~fm.  We are currently working to include more realistic temperature and time dependent parameters, and that may change these values. It will undoubtedly be useful to use hydrodynamic simulations as in refs.\ \cite{Young:2014pka,Nagai:2016wyx} to compute this quantity, although simulations in ref.\  \cite{Sharma:2011nj} suggest that millions of events are needed for sufficient numerical accuracy.

\begin{acknowledgments}
We are grateful to Rajendra Pokharel for collaboration in the early stages of this work. We thank Monika Sharma and Claude Pruneau for discussing the STAR data. We thank Victoria Drolshagen, Mauricio Martinez, Jaki Noronha-Hostler, Jorge Noronha, Scott Pratt, and Clint Young.  This work was supported in part by the U.S. NSF grant PHY-1207687. 
\end{acknowledgments}

\bibliography{ptDiffusion_References}

\end{document}